# CIC: Circular Image Compression

Honggui LI[1*], Sinan CHEN[2], Dingtai LI[3], Zhengyang ZHANG[4], Nahid MD LOKMAN HOSSAIN[5], Xinfeng XU[6], Yinlu QIN[7], Ruobing WANG[8], Maria TROCAN[9], Dimitri GALAYKO[10], Amara AMARA[11], Mohamad SAWAN[12,13]

[1,2,4,5,6,7,8]School of Information and Artificial Intelligence, Yangzhou University, Yangzhou 225127, China

[3]Shanghai Qigong Research Institute, Shanghai University of Traditional Chinese Medicine, Shanghai 201203, China

[9]LISITE Research Laboratory, Institut Supérieur d'Électronique de Paris, Paris 75006, France

[10]Laboratoire d'Informatique de Paris 6, Sorbonne University, Paris 75020, France

[11]International Campus (Hangzhou), Beijing University of Aeronautics and Astronautics, Hangzhou 310000, China

[12]School of Engineering, Westlake University, Hangzhou 310024, China

[13]Polystim Neurotech Laboratory, Polytechnique Montreal, Montreal H3T1J4, Canada

[1]hgli@yzu.edu.cn, [2]mz120240994@stu.yzu.edu.cn, [3]22025067@shutcm.edu.cn, [4]mz120250945@stu.yzu.edu.cn, [5]mh23083@stu.yzu.edu.cn, [6]241301123@stu.yzu.edu.cn, [7]241301115@stu.yzu.edu.cn, [8]241303321@stu.yzu.edu.cn, [9]maria.trocan@isep.fr, [10]dimitri.galayko@sorbonne-universite.fr, [11]amara@buaa.edu.cn, [12]sawan@westlake.edu.cn, [13]mohamad.sawan@polymtl.ca

[*]Corresponding Author

**Abstract**: Learned image compression (LIC) is currently the cutting-edge method. However, the inherent difference between testing and training images of LIC results in performance degradation to some extent. Especially for out-of-sample, out-of-distribution, or out-of-domain testing images, the performance of LIC degrades significantly. Classical LIC is a serial image compression (SIC) approach that utilizes an open-loop architecture with serial encoding and decoding units. Nevertheless, according to the principles of automatic control systems, a closed-loop architecture holds the potential to improve the dynamic and static performance of LIC. Therefore, a circular image compression (CIC) approach with closed-loop encoding and decoding elements is proposed to minimize the gap between testing and training images and upgrade the capability of LIC. The proposed CIC establishes a nonlinear loop equation and proves that steady-state error between reconstructed and original images is close to zero by Taylor series expansion. The proposed CIC method possesses the property of Post-Training and Plug-and-Play which can be built on any existing advanced SIC methods. Experimental results including rate-distortion curves on five public image compression datasets demonstrate that the proposed CIC outperforms eight competing state-of-the-art open-source SIC algorithms in reconstruction capacity. Experimental results further show that the proposed method is suitable for out-of-sample testing images with dark backgrounds, sharp edges, high contrast, grid shapes, or complex patterns.

# 1. INTRODUCTION

In the contemporary world of big data and generative artificial intelligence, a large amount of images are produced in various ways moment by moment [1]. Hence, it is necessary to efficiently compress images before transmission, storage, analysis, processing, recognition, and understanding [2-3]. Image compression methods can be divided into two categories: lossy and lossless image compression [4]. The former seeks for the balance between bitrate and distortion, the latter seeks for minimum bitrate without distortion. Image compression methods can also be partitioned into two categories: model-based and learning-based image compression [5]. The former develops to the full, and the latter is the front research direction and can be named as end-to-end deep learning-based image compression, learning-driven image compression, or learned image compression (LIC).

Because LIC is data-driven, deep neural network is firstly optimized by training image datasets and finally evaluated by testing image datasets. Training and testing image datasets hold similar but not identical characteristics. Deep neural network is optimal for training image datasets and is not always optimal for testing image datasets. Thus, the discrepancy between testing and training image datasets leads to performance degradation to some degree. Particularly for out-of-sample, out-of-distribution, or out-of-domain testing images, the performance of LIC dramatically degrades [6-8]. Hence, it is vitally important to improve the output of trained deep neural network based on testing image datasets to achieve ideal image reconstruction performance.

LIC usually utilizes an open-loop architecture with serial encoding and decoding units and can be named as serial image compression (SIC). Closed-loop architecture is widely adopted in automatic control systems to obtain extraordinary static and dynamic performance. According to the theory of automatic control, closed-loop architecture is superior to open-loop architecture in steady and transient states capability [9-10]. Therefore, circular image compression (CIC) method with closed-loop encoding and decoding elements is proposed to minimize the gap between training and testing image datasets and improve the reconstruction performance of LIC. It is worth mentioning that the proposed method is an extension of our previous work, circular image super-resolution, circular image compressive sensing, and circular compressed image super-resolution [11-13].

The proposed CIC is described by a nonlinear loop equation which is resolved by Taylor series expansion. Taylor series is an efficient tool for nonlinear analysis and has already been used for deep learning and LIC [14-15]. P. X. Wei et al propose a Taylor neural network with Taylor series approximation [14]. Y. E. Bao et al present a Taylor series expansion of sinusoidal functions based two-branch nonlinear transformation architecture to eliminate correlations from images [15].

In the present realm of large language models, there are many excellent pretrained SIC models which are openly and freely released on the Internet, such as GitHub and HuggingFace. They provide the opportunity to upgrade the performance of the pretrained SIC models in the way of Plug-and-Play and Post-Training. Actually, Plug-and-Play policy is widely employed in deep learning-based image restoration including the decoding of lossy image compression [16-21]. The proposed CIC can be built upon any existing advanced pretrained SIC models, enhancing their performance as a Plug-and-Play, post-training solution.

The core novelties of this paper are outlined as follows:

- closed-loop CIC framework with encoding and decoding elements;
- Plug-and-Play and Post-Training attribution grounded on any leading pretrained SIC models;
- nonlinear loop equation and complete mathematical proof of zero steady-state error with linear approximation of Taylor series expansion;
- huge performance boost in peak signal-to-noise ratio (PSNR), structural similarity (SSIM), and bits per sub pixel (BPSP); absolute and logic difference image blocks which demonstrate the extraordinary reconstruction capability.

The remainder of this paper is arranged as follows. The related work is reviewed in section 2, the theoretical fundamentals are elaborated in section 3, the evaluation experiment is conducted in section 4, and the summary and prospect are discussed in section 5.

## 2. RELATED WORK

With the swift progress of deep learning theory and technology, LIC methods continuously improve their performance [22-46]. These methods can be divided into two categories: In-Training and Post-Training based methods. The former enhances LIC performance during training procedure, the latter enhances LIC performance after training procedure or during testing procedure. Both In-Training and Post-Training methods focus on encoding-decoding network architectures, entropy models of latent representations, quantization policies, attention mechanisms, etc.

In-Training based LIC methods utilize some state-of-the-art generative models, such as diffusion model, flow model, autoregressive model, generative adversarial network (GAN), variance autoencoder (VAE), residual network (ResNet) based model, transformer-based model, convolutional neural network (CNN) based model, and so on [22-29]. Y. C. Bai et al propose a VAE and autoregressive model based deep lossy plus residual coding method for both lossless and near-lossless image compression [22]. R. H. Yang et al present a lossy image compression approach with conditional diffusion model [23]. Y. C. Bai et al also raise an end-to-end image compression algorithm with transformer-based model [24]. Z. B. Zhang et al put forward a decoupled framework-based image compression method that lets autoregressive model hold the capability of decoding in parallel [25]. D. Y. Zhang et al come up with a resolution field-based reciprocal pyramid network for scalable image compression [26]. N. D. Jr Guerin et al propose a VAE-based LIC method that dynamically adapts loss parameters to mitigate rate estimation issues and ensure precise target bitrate attainment [27]. W. C. Zhang et al present a semantically disentangled ultra-low bitrate LIC codec by synthesizing multiple neural computing techniques such as style GAN, inverse GAN mapping, and contrastive disentangled representation learning [28]. H. S. Fu et al raise a flexible discretized Gaussian-Laplacian-Logistic mixture model for the latent representations, which can adapt to different contents in different images and different regions of one image more accurately and efficiently [29].

Some In-Training based LIC methods concentrate on quantization strategies of latent representations [30-35]. Z. H. Duan et al put forward a lossy image compression approach with quantized hierarchical VAE [30]. A. Timur comes up with a vector quantized VAE for image compression [31]. S. L. Cai et al propose a flow model based invertible continuous codec for high-fidelity variable-bitrate image compression to avoid the usage of a set of different models for compressing images at different rates [32]. H. S. Fu et al present an asymmetric LIC algorithm with multi-scale residual block, importance scaling, and post-quantization filtering [33]. G. Zhang et al raise an enhanced quantified local implicit neural representation for image compression by enhancing the utilization of local relationships of implicit neural representation and narrow the quantization gap between training and encoding [34]. J. Y. Guo et al put forward a new LIC framework that aims to learn one single network to support variable bitrate coding under various computational complexity levels [35].

Some In-Training based LIC methods are concerned with attention mechanisms [36-38]. Z. Y. Jiang et al come up with a novel image compression autoencoder based on the local-global joint attention mechanism [36]. B. Li et al propose LIC approach via neighborhood-based attention optimization and context modeling with multi-scale guiding [37]. Z. S. Tang et al present an end-to-end image compression method integrating graph attention and asymmetric CNN [38].

Pots-Training based LIC methods strengthen the performance of pretrained LIC models [39-42]. J. Q. Shi et al adopt a Plug-and-Play rate-distortion optimized Post-Training quantization to process pretrained, off-the-shelf LIC models and minimize quantization-induced error of model parameters [39]. Z. H. Duan et al raise a quantization-aware ResNet VAE for lossy image compression with test-time quantization and quantization-aware training [40]. S. H. Li et al put forward a progressive LIC algorithm with dead-zone quantizers on the latent representation which is successfully incorporated into existing pretrained fixed-rate models without re-training [41]. H. Son et al come up with an enhanced standard compatible image compression framework to fuse learnable codecs, postprocessing networks, and compact representation networks [42].

Some Pots-Training based LIC methods directly enhance the quality of decoded images of LIC [43-46]. J. F. Li et al propose a recurrent convolution network for blind image compression artifact reduction in industrial IoT systems [43]. L. Ma et al present a sensitivity decouple learning approach for image compression artifacts reduction which decouples the intrinsic image attributes into compression-insensitive features for high-level semantics and compression-sensitive features for low-level cues [44]. J. H. Hu et al raise a ResNet for image compression artifact reduction [45]. H. G. Chen et al put forward a deep CNN for JPEG image compression artifacts reduction [46].

Some In-Training and Post-Training based LIC methods attempt to resolve the problem of out-of-sample, out-of-distribution, or out-of-domain testing images [7-8]. S. H. Li et al come up with a Post-Training pruning method based on the admissible range and in-distribution region to automatically remove the out-of-distribution channels for LIC [7]. K. Tsubota et al propose a content-adaptive optimization framework for universal LIC which adapts a pretrained compression model to each target image during testing for addressing the domain gap between pretraining and post-testing [8].

In summary, this paper presents a Post-Training based lossy LIC method, CIC, to minimize the discrimination between testing and training image datasets and promote the performance of image reconstruction.

## 3. THEORY

### *3.1. Technical Lexicon*

Table 1 gathers the technical abbreviations and mathematical notations employed in this paper.

Table 1 AGGREGATION OF ABBREVIATIONS & NOTATIONS

| **Abbreviation & Notation** | **Meaning** |
|---|---|
| EN / DE | Encoding / Decoding |
| SIC / CIC | Serial / Circular Image Compression |
| NF | Nonlinear Function |
| D / d | Original / Encoded Image Dimension |
| W / H / C | Image Width / Height / Channels |
| $\mathbf{f}$ / $\mathbf{f}_e$ / $\mathbf{f}_d$ | Original / Encoded / Decoded Image |
| $\mathbf{f}_r$ / $\mathbf{f}_c$ | Residual Term / Control Term |
| $\mathbf{f}_a$ / $\mathbf{f}_l$ | Absolute / Logical Difference Image |
| $\mathbf{f}_t$ / $\mathbf{f}_r$ | Testing / Reference Difference Image |
| $\mathbf{o}$ / $\mathbf{O}$ | First-Order Term / Higher-Order Term |
| $\mathbf{\Lambda}$ / $\mathbf{U}$ | Coefficient Matrix |
| t / $\mathbf{r}$ | Time / Reconstruction Error |
| N / η | Iteration Number / Iteration Constant |
| LIC | Learned Image Compression |
| VAE | Variational Auto-Encoders |
| DLPR | Deep Lossy Plus Residual Coding |
| CDC | Conditional Diffusion Compression |
| ICT | Image Compression with Transformers |
| QRVAE | Quantized ResNet Variational Auto-Encoders |
| VQVAE | Vector Quantized Variational Auto-Encoders |
| CGIC | Controllable Generative Image Compression |
| TACO | Text-Adaptive Compression |
| LBIC | Learned Block-based Image Compression |
| PSNR | Peak Signal-to-Noise Ratio |
| SSIM | Structural Similarity |
| BPSP | Bits Per Sub Pixel |

*3.2. Conceptual Infrastructure*

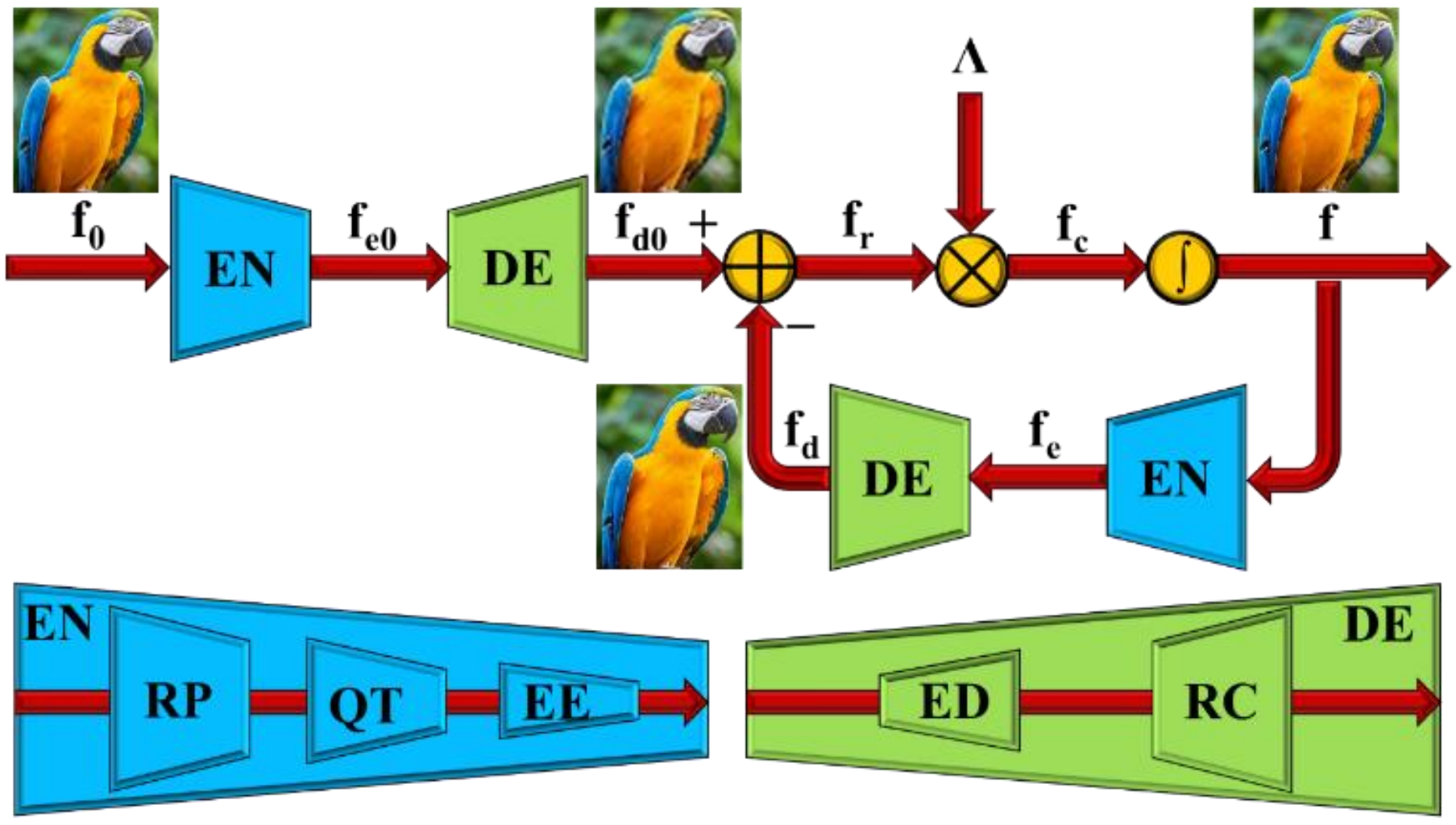

**Fig. 1.** The conceptual infrastructure of CIC.

Figure 1 illustrates the proposed conceptual infrastructure of CIC. It is a combination of open-loop and closed-loop architectures. It comprises two branches: the left branch and the right branch. The left branch is the classical open-loop SIC framework and can also be designated as linked or cascade image compression. The right branch is the closed-loop CIC framework and can also be designated as ringed or cycle image compression. SIC includes two elements: encoding (EN) and decoding (DE). The EN element compresses the original image to the encoded image and is composed of representation (RP), quantization (QT), and entropy coding (EC). The DE element decompresses the encoded image to the decoded image and is composed of entropy decoding (ED) and reconstruction (RC). CIC embraces five subparts: EN, DE, summator, multiplier, and integrator. The EN and DE subparts of CIC coincide with those of SIC. The summator imports negative feedback into CIC. The multiplier, integrator, and summator implement the traditional proportion-integration control. The input of the proposed infrastructure is original image $\mathbf{f}_0$ and the output is expected reconstructed image $\mathbf{f}$. The proposed CIC can achieve better reconstruction images than those of the traditional SIC.

The proposed CIC holds the property of Plug-and-Play and can be built on any existing advanced SIC methods. That is to say, the proposed CIC can take advantage of the EN and DE units of any pretrained SIC models.

The SIC can be depicted by the following mathematical expressions:

$$\mathbf{f}_{e0}=\mathrm{EN}\left(\mathbf{f}_0\right),\mathbf{f}_{d0}=\mathrm{DE}\left(\mathbf{f}_{e0}\right),\mathbf{f}_0,\mathbf{f}_{d0}\in \mathrm{R}^{\mathrm{D}};\mathbf{f}_{e0}\in \mathrm{R}^{\mathrm{d}}, \quad (1)$$

where: EN denotes the encoder and is specified in the classical SIC; DE denotes the decoder and is also specified in the classical SIC; $\mathbf{f}_0$ denotes the original image; $\mathbf{f}_{e0}$ denotes the original encoded image after EN and is the true input of the proposed infrastructure; $\mathbf{f}_{d0}$ denotes the original decoded image after DE; D denotes the dimension of $\mathbf{f}_0$ and $\mathbf{f}_{d0}$; d denotes the dimension of $\mathbf{f}_{e0}$.

According to Equation (1), the nonlinear function (NF) of SIC from the input to the output can be portrayed by the following mathematical formulas:

$$\mathbf{f}_{d0}=\mathrm{DE}\left(\mathrm{EN}\left(\mathbf{f}_0\right)\right)=\mathrm{NF}\left(\mathbf{f}_0\right),\mathrm{NF}\left(\cdot\right)=\mathrm{DE}\left(\mathrm{EN}\left(\cdot\right)\right),\mathrm{NF}\left(\cdot\right)\in \mathrm{R}^{\mathrm{D}}. \quad (2)$$

The CIC can be described by the following mathematical equations:

$$\begin{gathered}\mathbf{f}_e\left(t\right)=\mathrm{EN}\left(\mathbf{f}\left(t\right)\right),\mathbf{f}_d\left(t\right)=\mathrm{DE}\left(\mathbf{f}_e\left(t\right)\right),\mathbf{f}_r\left(t\right)=\mathbf{f}_{d0}-\mathbf{f}_d\left(t\right),\mathbf{f}_c\left(t\right)=\mathbf{\Lambda}\left(t\right)\mathbf{f}_r\left(t\right),\mathbf{f}\left(t\right)=\mathbf{f}\left(0\right)+\int_0^t\mathbf{f}_c\left(t\right)dt \\ \mathbf{\Lambda}\left(t\right)=\mathrm{diag}\left(\mathbf{\Lambda}_{11}\left(t\right)\quad \mathbf{\Lambda}_{22}\left(t\right)\quad \cdots \quad \mathbf{\Lambda}_{DD}\left(t\right)\right),t\in \mathrm{R};\mathbf{f}\left(t\right),\mathbf{f}_d\left(t\right),\mathbf{f}_r\left(t\right),\mathbf{f}_c\left(t\right)\in \mathrm{R}^{\mathrm{D}};\mathbf{f}_e\left(t\right)\in \mathrm{R}^{\mathrm{d}};\mathbf{\Lambda}\left(t\right)\in \mathrm{R}^{\mathrm{D}\times \mathrm{D}}\end{gathered}, \quad (3)$$

where: t is the time; $\mathbf{f}(t)$ is the expected reconstructed image at time t and is the output of the proposed infrastructure; $\mathbf{f}_e(t)$ is the expected encoded image of $\mathbf{f}(t)$ at time t; $\mathbf{f}_d(t)$ is the expected decoded image of $\mathbf{f}_e(t)$ at time t; $\mathbf{f}_r(t)$ is the residual term at time t; $\mathbf{f}_c(t)$ is the control term at time t; $\mathbf{f}(0)$ is the initial value of $\mathbf{f}(t)$ at time 0 and bears equality to zero vector, random vector, or $\mathbf{f}_{d0}$; $\mathbf{\Lambda}(t)$ is the diagonal matrix of multiplication coefficient at time t; $\mathbf{\Lambda}_{ii}(t)$ is the i-th principal diagonal item of $\mathbf{\Lambda}(t)$; diag($\cdot$) is a diagonal matrix.

In accordance with the closed-loop control theory, when the proposed framework is steady, the residual term $\mathbf{f}_r(t)$ is close to zero, then the expected decoded image $\mathbf{f}_d(t)$ is close to the original decoded image $\mathbf{f}_{d0}$, and finally the expected reconstructed image $\mathbf{f}(t)$ is close to the original image $\mathbf{f}_0$. Therefore, ideal reconstruction image is achieved. The related mathematical proof is provided in following subsection of steady-state error.

The proposed infrastructure in Fig. 1 is straightforward, simple, and easy to be implemented because the multiplier and summator are apt to be achieved, the integrator can be realized by a summator, and the EN and DE units can make use of any existing SIC algorithms. However, it should be mentioned that the proposed CIC outperforms classical SIC at the cost of doubling computation load. The proposed CIC is appropriate for the applications with abundant computing resources, not for the applications with constrained computing resources.

### *3.3. Loop Equation*

As derived from Equations (2) and (3), the nonlinear loop equation of CIC can be depicted by the following mathematical expressions:

$$\mathbf{f}(t)=\mathbf{f}(0)+\int_0^t \boldsymbol{\Lambda}(t)\left(\mathbf{f}_{d0}-\mathrm{DE}\left(\mathrm{EN}\left(\mathbf{f}(t)\right)\right)\right)dt,\mathbf{f}(t)=\mathbf{f}(0)+\int_0^t \boldsymbol{\Lambda}(t)\left(\mathbf{f}_{d0}-\mathrm{NF}\left(\mathbf{f}(t)\right)\right)dt\ . \quad (4)$$

*3.4. Nonlinear Function*

The nonlinear function NF($\mathbf{f}$(t)) in Equation (4) can be expressed via the Taylor series at $\mathbf{f}_0$ by the following mathematical equalities:

$$\mathrm{NF}\left(\mathbf{f}(t)\right)=\mathrm{NF}\left(\mathbf{f}_0\right)+\mathbf{o}(t)+\mathbf{O}(t),\mathbf{o}_i(t)=\sum_{j=1}^{D}\left(\mathbf{f}_j(t)-\mathbf{f}_{0j}\right)\left.\frac{\partial \mathrm{NF}_i\left(\mathbf{f}(t)\right)}{\partial \mathbf{f}_j(t)}\right|_{\mathbf{f}(t)=\mathbf{f}_0},\mathbf{o}(t),\mathbf{O}(t)\in R^D;i,j=1,2,\cdots,D\ , \quad (5)$$

where: NF($\mathbf{f}_0$) represents the constant term of NF($\mathbf{f}$(t)) at $\mathbf{f}_0$; $\mathbf{o}$(t) represents the first-order term of NF($\mathbf{f}$(t)) at time t; $\mathbf{O}$(t) represents the second-order and higher-order term of NF($\mathbf{f}$(t)) at time t; $\mathbf{o}_i$(t) represents the i-th item of $\mathbf{o}$(t) where the partial derivatives can be calculated by difference operation; $\mathbf{f}_j$(t) represents the j-th item of $\mathbf{f}$(t); $\mathbf{f}_{0j}$ represents the j-th item of $\mathbf{f}_0$; $\mathrm{NF}_i$($\mathbf{f}$(t)) represents the i-th item of NF($\mathbf{f}$(t)).

For the purpose of subsequent theoretical analysis, after abandoning the second-order and higher-order term $\mathbf{O}$(t), the NF($\mathbf{f}$(t)) can be approximately given by the following linear mapping:

$$\mathrm{NF}\left(\mathbf{f}(t)\right)\approx \mathrm{NF}\left(\mathbf{f}_0\right)+\mathbf{o}(t)\ . \quad (6)$$

It is remarkable that this linear approximation is local at $\mathbf{f}_0$ and the whole loop equation is still nonlinear. In addition, the higher-order term is smaller enough compared with linear term and can be omitted.

The linear term $\mathbf{o}_i$(t) can further take the approximate form of the following mathematical equations:

$$\mathbf{o}_i(t)\approx\left(\mathbf{f}_i(t)-\mathbf{f}_{0i}\right)\left.\frac{\partial \mathrm{NF}_i\left(\mathbf{f}\right)}{\partial \mathbf{f}_i(t)}\right|_{\mathbf{f}_i(t)=\mathbf{f}_{0i}},i=1,2,\cdots,D\ . \quad (7)$$

In light of Equation (7), the linear term $\mathbf{o}$(t) can be adapted by the following formulas:

$$\mathbf{o}(t)\approx\mathbf{U}(t)\left(\mathbf{f}(t)-\mathbf{x}_0\right),\mathbf{U}(t)=\mathrm{diag}\left(\mathbf{U}_{11}(t)\quad \mathbf{U}_{22}(t)\quad \cdots \quad \mathbf{U}_{DD}(t)\right),\mathbf{U}_{ii}(t)=\left.\frac{\partial \mathrm{NF}_i\left(\mathbf{f}(t)\right)}{\partial \mathbf{f}_i(t)}\right|_{\mathbf{f}(t)=\mathbf{f}_0}, \quad (8)$$
$$\mathbf{U}(t)\in R^{D\times D};i=1,2,\cdots,D$$

where: $\mathbf{U}$(t) denotes the coefficient matrix at time t; $\mathbf{U}_{ii}$(t) denotes the i-th principal diagonal element of $\mathbf{U}$(t).

*3.5. Steady-State Error*

The reconstruction error between the expected reconstruction image $\mathbf{f}$(t) and the original image $\mathbf{f}_0$ can be depicted by the following mathematical expression:

$$\mathbf{r}(t)=\mathbf{f}(t)-\mathbf{f}_0,\mathbf{r}(t)\in R^D\ . \quad (9)$$

When t is close to infinite and the proposed framework is in steady-state, the reconstruction error r(∞) is named as steady-state error.

Pursuant to Equation (4), the expected reconstruction image **f**(t) at time t+Δt can be portrayed by the following mathematical formula:

$$\mathbf{f}(t+\Delta t)=\mathbf{f}(t)+\int_{t}^{t+\Delta t}\mathbf{\Lambda}(t)\Big(\mathbf{f}_{d0}-\mathrm{NF}\big(\mathbf{f}(t)\big)\Big)dt, \text{s.t. } \Delta t>0 . \ (10)$$

Deducting $\mathbf{f}_0$ from either side of Equation (10), the following mathematical expression can be gained:

$$\mathbf{f}(t+\Delta t)-\mathbf{f}_0=\mathbf{f}(t)-\mathbf{f}_0+\int_{t}^{t+\Delta t}\mathbf{\Lambda}(t)\Big(\mathbf{f}_{d0}-\mathrm{NF}\big(\mathbf{f}(t)\big)\Big)dt . \ (11)$$

By virtual of Equation (9), the following mathematical equations can be acquired:

$$\mathbf{r}(t+\Delta t)=\mathbf{r}(t)+\int_{t}^{t+\Delta t}\mathbf{\Lambda}(t)\Big(\mathbf{f}_{d0}-\mathrm{NF}\big(\mathbf{f}(t)\big)\Big)dt, \mathbf{r}(t+\Delta t)=\mathbf{f}(t+\Delta t)-\mathbf{f}_0 , \ (12)$$

where **r**(t+Δt) means the error vector at time t+Δt.

If Δt approaches zero, Equation (12) can be expressed approximately by the following mathematical equality:

$$\mathbf{r}(t+\Delta t)\approx\mathbf{r}(t)+\mathbf{\Lambda}(t)\Big(\mathbf{f}_{d0}-\mathrm{NF}\big(\mathbf{f}(t)\big)\Big)\Delta t, \text{s.t. } \Delta t\to 0 . \ (13)$$

According to Equations (2), (6), (8), and (9), the following mathematical formulas can be achieved:

$$\begin{aligned}\mathbf{r}(t+\Delta t)&\approx\mathbf{r}(t)-\mathbf{\Lambda}(t)\mathbf{o}(t)\Delta t\approx\mathbf{r}(t)-\mathbf{\Lambda}(t)\mathbf{U}(t)\big(\mathbf{f}(t)-\mathbf{f}_0\big)\Delta t=\mathbf{r}(t)-\mathbf{\Lambda}(t)\mathbf{U}(t)\mathbf{r}(t)\Delta t\\&=\big(\mathbf{I}-\mathbf{\Lambda}(t)\mathbf{U}(t)\Delta t\big)\mathbf{r}(t)=\big(\mathbf{I}-\Delta t\mathbf{\Lambda}(t)\mathbf{U}(t)\big)\mathbf{r}(t)\end{aligned} ,(14)$$

where **I** represents the unit matrix.

Calculating norm in the two sides of Equation (14), the following mathematical inequality can be gained:

$$\left\|\mathbf{r}(t+\Delta t)\right\|_2\le\left\|\mathbf{I}-\Delta t\mathbf{\Lambda}(t)\mathbf{U}(t)\right\|_F\cdot\left\|\mathbf{r}(t)\right\|_2 , \ (15)$$

where: subscript 2 signifies 2-norm; subscript F signifies Frobenius-norm.

It is invariably possible to select proper **Λ**(t) and **U**(t) to meet the following inequation:

$$\left\|\mathbf{I}-\Delta t\mathbf{\Lambda}(t)\mathbf{U}(t)\right\|_F<1 . \ (16)$$

For instance, **Λ**(t) and **U**(t) are respectively commensurate to a unit matrix:

$$\mathbf{\Lambda}(t)=\eta\mathbf{I}, \mathbf{U}(t)=\mu(t)\mathbf{I}, \mu(t)=\frac{1}{D}\sum_{i=1}^{D}\mathbf{U}_{ii}(t)=\frac{1}{D}\sum_{i=1}^{D}\left.\frac{\partial \mathrm{NF}_i\big(\mathbf{f}(t)\big)}{\partial \mathbf{f}_i(t)}\right|_{\mathbf{f}(t)=\mathbf{f}_0} , \ (17)$$

where: η denotes a constant; μ(t) denotes the average of $\mathbf{U}_{ii}(t)$ in Equation (8).

In accordance with Equation (17), it is consistently possible to seek out a suitable η to satisfy Inequality (16):

$$\left\|\mathbf{I}-\Delta t\mathbf{\Lambda}(t)\mathbf{U}(t)\right\|_F=\left\|\mathbf{I}-\Delta t\cdot\eta\mathbf{I}\cdot u(t)\mathbf{I}\right\|_F=\left\|(1-\Delta t\cdot\eta\cdot u(t))\mathbf{I}\right\|_F=\left|1-\Delta t\cdot u(t)\cdot\eta\right|\cdot\left\|\mathbf{I}\right\|_F=\left|1-\Delta t\cdot u(t)\cdot\eta\right|\cdot\sqrt{D}<1$$
$$\Rightarrow\begin{cases}\dfrac{1-\frac{1}{\sqrt{D}}}{\Delta t\cdot u(t)}<\eta<\dfrac{1+\frac{1}{\sqrt{D}}}{\Delta t\cdot u(t)}, u(t)>0\\[2ex]\dfrac{1+\frac{1}{\sqrt{D}}}{\Delta t\cdot u(t)}<\eta<\dfrac{1-\frac{1}{\sqrt{D}}}{\Delta t\cdot u(t)}, u(t)<0\end{cases}. \quad (18)$$

Based on Equations (15) and (16), the following mathematical inequation can be acquired:

$$\left\|\mathbf{r}(t+\Delta t)\right\|_2<\left\|\mathbf{r}(t)\right\|_2. \quad (19)$$

As derived from Inequality (19), if time t is approximate to infinite in steady-state, the reconstruction error, steady-state error, is close to zero, and **f**(t) is nearly equal to $\mathbf{f}_0$. It can be portrayed by the following mathematical limit:

$$\left\|\mathbf{r}(\infty)\right\|_2=\lim_{t\to\infty}\left\|\mathbf{r}(t)\right\|_2=\lim_{t\to\infty}\left\|\mathbf{f}(t)-\mathbf{f}_0\right\|_2=0\Rightarrow\mathbf{r}(\infty)=\lim_{t\to\infty}\mathbf{r}(t)=\lim_{t\to\infty}(\mathbf{f}(t)-\mathbf{f}_0)=0\Rightarrow\lim_{t\to\infty}\mathbf{f}(t)=\mathbf{f}_0. \quad (20)$$

Hence, the proposed CIC can obtain the perfect reconstruction image **f**(t) which is approximate to the original image $\mathbf{f}_0$.

### *3.6. Algorithm Formulation*

Algorithm 1 describes the proposed CIC algorithm, where N is the total number of iterations.

**Algorithm 1: CIC.**

**Input: $f_0$**

**Initialization:**

  **n = 1, $f_{d0}$ = NF($f_0$), f = 0, $f_{d0}$, or random vector**

**while n <= N**

  **update f in accordance with equations (4), (6), (11) and (17)**

  **n = n+1**

**end**

**Output: f**

## 4. EXPERIMENT

### *4.1. Simulation Setup*

Five public image compression datasets, Kodak, CLIC2021 Test, CLIC2021 Validation, CLIC2022 Validation, and CLIC2024 Validation, are utilized to evaluate the performance of LIC methods. Table 2 enumerates these image datasets including the total number of images, the resolution of images, and the web link of image datasets.

Table 2 IMAGE DATASETS FOR SIMULATION

| Dataset | Detail | |
|---|---|---|
| Kodak | Number | 24 |
| | Resolution | 768×512 |
| | Link | https://www.kaggle.com/datasets/sherylmehta/kodak-dataset |
| CLIC2021 Test | Number | 60 |
| | Resolution | 751×500 ~ 2048×1400 |
| | Link | https://clic.compression.cc/2021/tasks/index.html |
| CLIC2021 Validation | Number | 41 |
| | Resolution | 512×384 ~ 2048×1370 |
| | Link | https://clic.compression.cc/2021/tasks/index.html |
| CLIC2022 Validation | Number | 30 |
| | Resolution | 1151×2048 ~ 2048×2048 |
| | Link | https://clic.compression.cc/2022/ |
| CLIC2024 Validation | Number | 30 |
| | Resolution | 1152×2048 ~ 2048×2048 |
| | Link | https://compression.cc/tasks/ |

Eight competing open-source methods of SIC, Deep Lossy Plus Residual Coding (DLPR) [22], Conditional Diffusion Compression (CDC) [23], Image Compression with Transformers (ICT) [24], Quantized ResNet VAE (QRVAE) [30], Vector Quantized VAE (VQVAE) [31], Controllable Generative Image Compression (CGIC) [47] , Text-Adaptive Compression (TACO) [48], and Learned Block-based Image Compression (LBIC) [49], are adopted for comparison with the proposed CIC method. Table 3 lists these methods including algorithm names and web links. Because the proposed CIC method has the property of Plug-and-Play and Post-Training, eight CIC versions of the competing methods, circular DLPR (CDLPR), circular CDC (CCDC), circular ICT (CICT), circular QRVAE (CQRVAE), circular VQVAE (CVQVAE), circular CGIC (CCGIC), circular TACO (CTACO), and circular LBIC (CLBIC) are presented for comparison.

Table 3 COMPETING METHODS

| Method | Link |
|---|---|
| DLPR | https://github.com/BYchao100/Deep-Lossy-Plus-Residual-Coding |
| CDC | https://github.com/buggyyang/CDC_compression |
| ICT | https://github.com/BYchao100/Towards-Image-Compression-and-Analysis-with-Transformers |
| QRVAE | https://github.com/duanzhiihao/qres-vae |
| VQVAE | https://github.com/TimeEscaper/vq-vic |
| CGIC | https://github.com/lianqi1008/Control-GIC |
| TACO | https://github.com/effl-lab/TACO |
| LBIC | https://github.com/kamisli-icpl/Learned-block-based-image-compression |

Four simulation experiments, quantization parameter, reconstruction performance, out-of-sample, and up-to-date algorithms are designed to compare the performance between the conventional SIC and the proposed CIC.

Three performance metrics, PSNR, SSIM, and BPSP, are selected to assess the capability of SIC and CIC. For reconstruction image $\mathbf{f}$ and original image $\mathbf{f}_0$, the computing methods of PSNR, SSIM, and BPSP are depicted by the following mathematical expressions:

$$\mathrm{PSNR}\left(\mathbf{f},\mathbf{f}_0\right)=10\lg\left(\frac{255^2}{\frac{1}{\mathrm{D}}\sum_{\mathrm{i}=1}^{\mathrm{D}}\left(\mathbf{f}_\mathrm{i}-\mathbf{f}_{0\mathrm{i}}\right)^2}\right), (21)$$

$$\mathrm{SSIM}\left(\mathbf{f},\mathbf{f}_0\right)=\frac{\left(2\mu_\mathbf{f}\mu_{\mathbf{f}_0}+\left(0.01\times255\right)^2\right)\left(2\sigma_{\mathbf{ff}_0}+\left(0.03\times255\right)^2\right)}{\left(\mu_\mathbf{f}^2+\mu_{\mathbf{f}_0}^2+\left(0.01\times255\right)^2\right)\left(\sigma_\mathbf{f}^2+\sigma_{\mathbf{f}_0}^2+\left(0.03\times255\right)^2\right)}, (22)$$

$$\mathrm{BPSP}=\frac{\mathrm{B}}{\mathrm{S}\times\mathrm{W}\times\mathrm{H}\times\mathrm{C}}, (23)$$

where: $\mathbf{f}_\mathrm{i}$ denotes the i-th element of $\mathbf{f}$; $\mathbf{f}_{0\mathrm{i}}$ denotes the i-th element of $\mathbf{f}_0$; $\mu_\mathbf{f}$ denotes the element-wised mean of $\mathbf{f}$; $\mu_{\mathbf{f}0}$ denotes the element-wised mean of $\mathbf{f}_0$; $\sigma_\mathbf{f}$ denotes the element-wised standard deviation of $\mathbf{f}$; $\sigma_{\mathbf{f}0}$ denotes the element-wised standard deviation of $\mathbf{f}_0$; $\sigma_{\mathbf{ff}0}$ denotes the element-wised covariance of $\mathbf{f}$ and $\mathbf{f}_0$; B denotes the total number of bits for an image; S denotes the total number of bits for a subpixel; W denotes the width of an image; H denotes the height of an image; C denotes the total number of channels of an image.

For the purpose of comparing the difference between the reconstruction images of SIC and CIC, absolute difference image and logic difference image are defined by the following mathematical formulas:

$$\begin{gathered}\mathbf{f}_\mathrm{a}\left(\mathrm{i},\mathrm{j}\right)=\left|\mathbf{f}_\mathrm{t}\left(\mathrm{i},\mathrm{j}\right)-\mathbf{f}_\mathrm{r}\left(\mathrm{i},\mathrm{j}\right)\right|\\ \mathbf{f}_\mathrm{l}\left(\mathrm{i},\mathrm{j}\right)=\begin{cases}1,\left|\mathbf{f}_\mathrm{t}\left(\mathrm{i},\mathrm{j}\right)-\mathbf{f}_\mathrm{r}\left(\mathrm{i},\mathrm{j}\right)\right|>\mathrm{T}\\ 0,\left|\mathbf{f}_\mathrm{t}\left(\mathrm{i},\mathrm{j}\right)-\mathbf{f}_\mathrm{r}\left(\mathrm{i},\mathrm{j}\right)\right|\le\mathrm{T}\end{cases}, (24)\\ \mathbf{f}_\mathrm{a},\mathbf{f}_\mathrm{l},\mathbf{f}_\mathrm{t},\mathbf{f}_\mathrm{r}\in\mathrm{R}^{\mathrm{H}\times\mathrm{W}};\mathrm{i},\mathrm{j}\in Z;\mathrm{T}\in\mathrm{R}\end{gathered}$$

where: $\mathbf{f}_\mathrm{a}$ means the absolute difference image; $\mathbf{f}_\mathrm{l}$ means the logic difference image; $\mathbf{f}$ means the testing image; $\mathbf{f}_\mathrm{r}$ means the reference image; i means the row index of an image; j means the column index of an image; T means a threshold.

The simulation hardware testbeds contain Intel CPU and Nvidia GPU. The simulation software testbeds contain Microsoft GitHub, Google TensorFlow and COLAB, FaceBook PyTorch, JetBrains PyCharm, and MathWorks MATLAB running on Windows or Linux operating systems. Table 4 itemizes comprehensive hardware and software specifications.

Table 4 COMPREHENSIVE HARDWARE AND SOFTWARE SPECIFICATIONS

| **Hardware Specifications** | |
|---|---|
| Entry | Setting |
| CPU Category | Intel Core i7 |
| CPU Memory | 16GB |
| GPU Category | NVIDIA Ampere A100 |
| GPU Memory | 32GB |
| **Software Specifications** | |
| Entry | Setting |
| Input/Output Channels | 3/3 |
| Batch Size | 1 |
| Block Size | 64×64 |
| Iteration Number (N) | 1~10 |
| Iteration Constant (η) | –1~+1 |

### *4.2. Experimental Results*

#### *4.2.1 Experimental Results of Quantization Parameter*

This experiment is designed to investigate the relationship between the performance of image compression and the quantization parameter. This experiment focuses on the DLPR and CDLPR algorithms and the single image, kodim01.png, of Kodak image dataset. Figure 2 shows that PSNR decreases while quantization parameter $\tau$ increases and CDLPR outperforms DLPR in PSNR. The PSNR increment $\Delta$ goes up first, then reaches a peak value at $\tau$=7, and finally goes down. Figure 3 also displays that SSIM decreases while quantization parameter $\tau$ increases and CDLPR exceeds DLPR in SSIM. The SSIM increment $\Delta$ rises first, then arrives at a crest value at $\tau$=7, and finally falls. Figure 4 further illustrates that BPSP decreases while quantization parameter $\tau$ increases and CDLPR surpasses DLPR in BPSP. The BPSP decrement $\Delta$ monotonically declines. A quantization parameter with maximum PSNR and SSIM increments will be chosen in following experiments. In addition, Figures 2-4 show that PSNR, SSIM, and BPSP of CDPLR are robust for varying compression rates because BPSP is inversely proportional to compression rate.

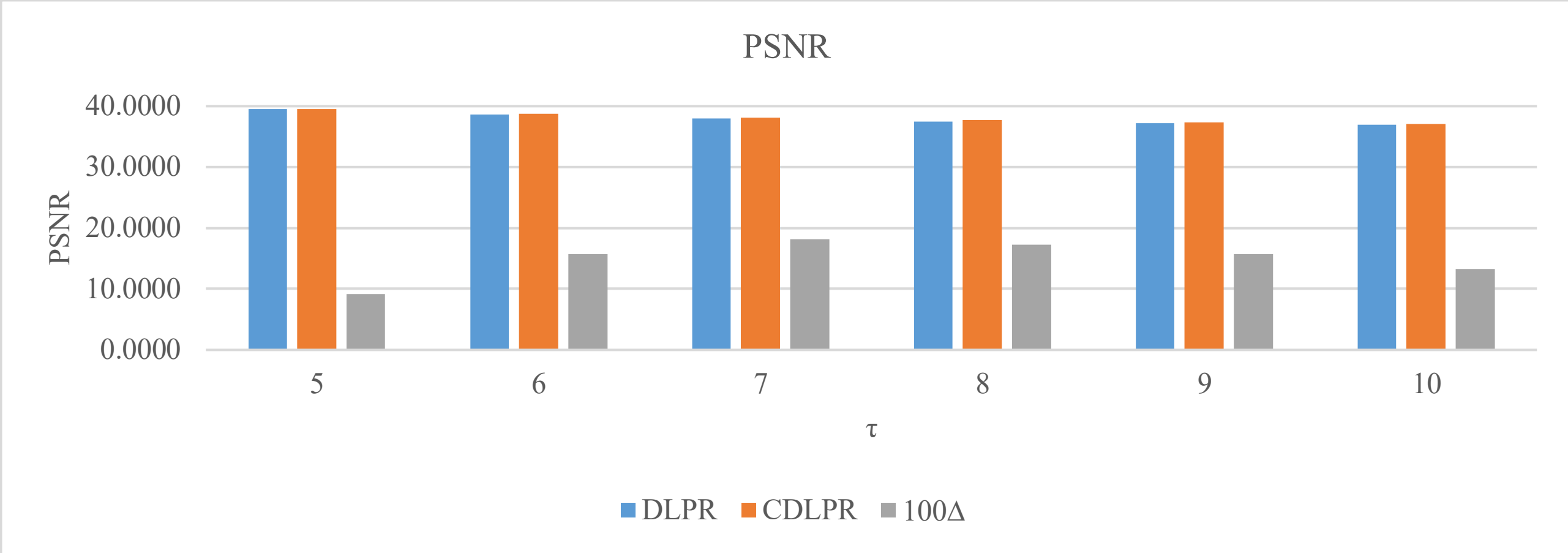

**Fig. 2.** PSNR of DLPR and CDLPR on single image of Kodak image dataset.

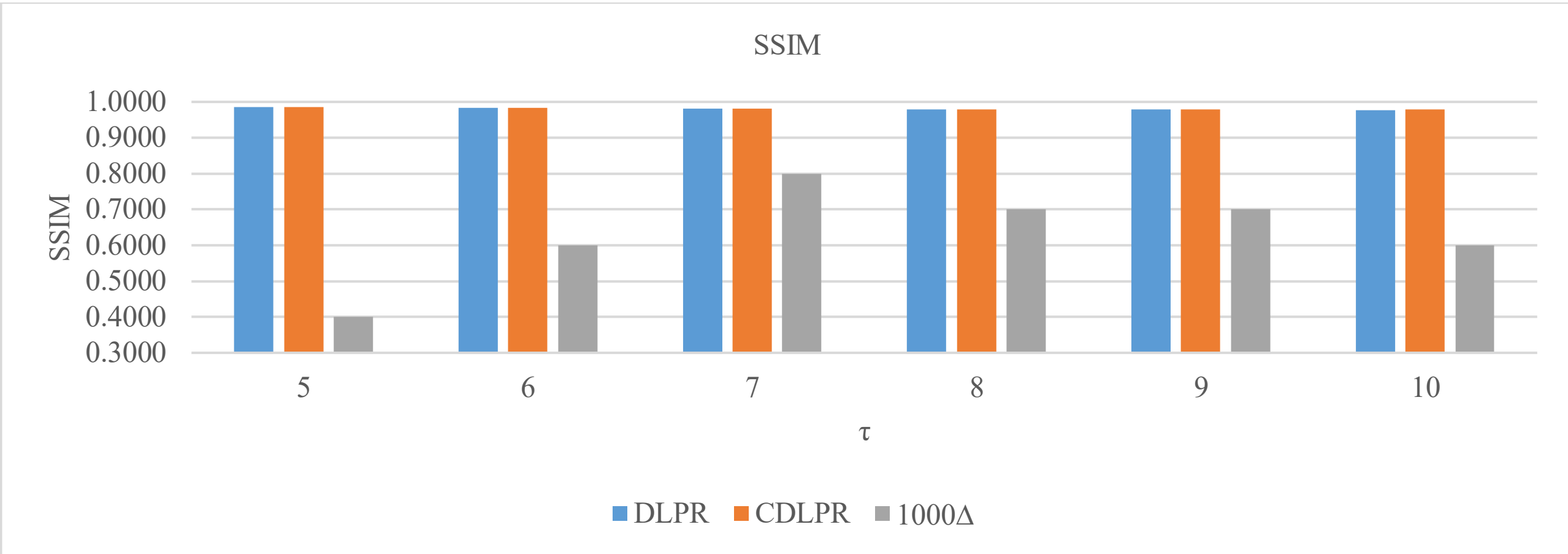

**Fig. 3.** SSIM of DLPR and CDLPR on single image of Kodak image dataset.

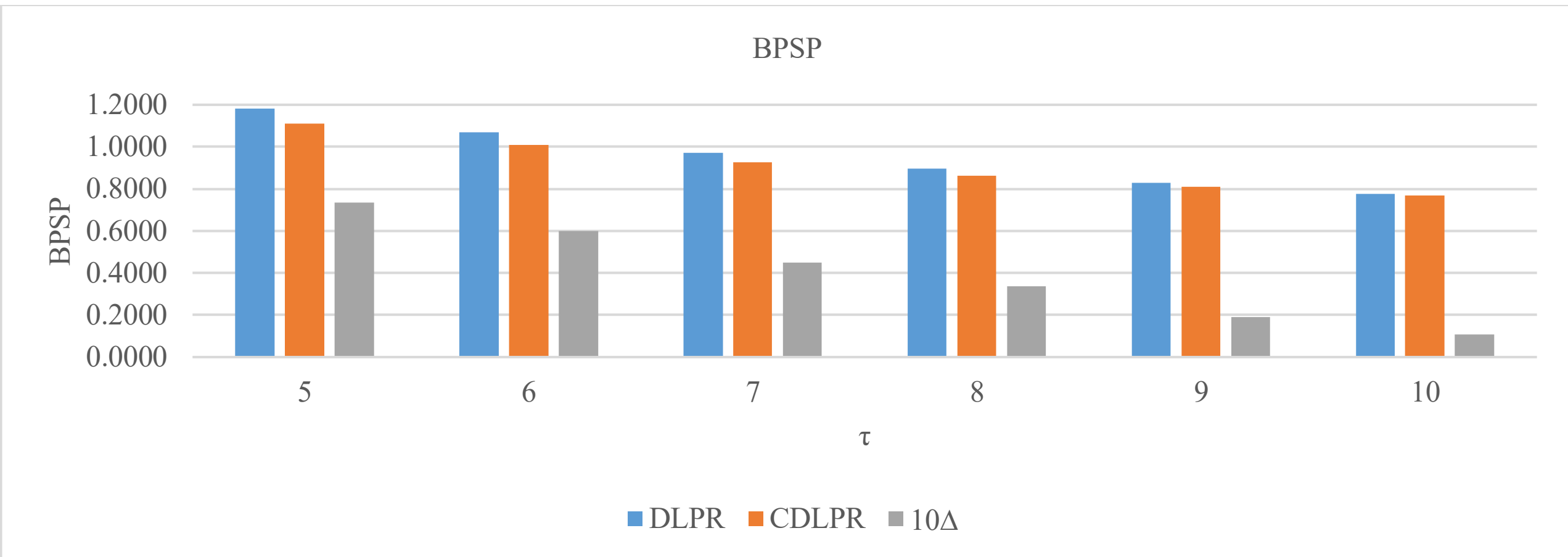

**Fig. 4.** BPSP of DLPR and CDLPR on single image of Kodak image dataset.

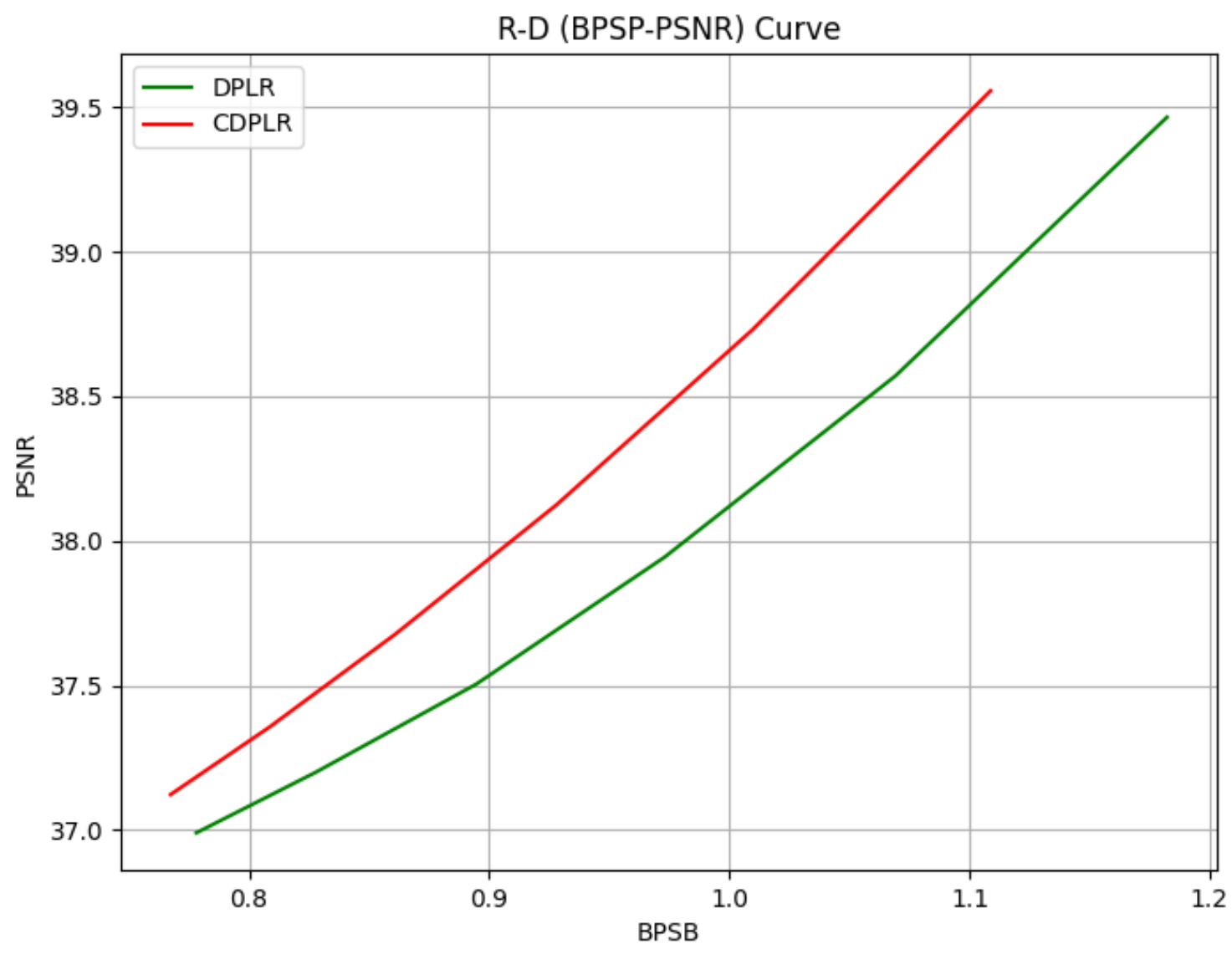

**Fig. 5.** R-D (BPSP-PSNR) curve of DLPR and CDLPR on single image of Kodak image dataset.

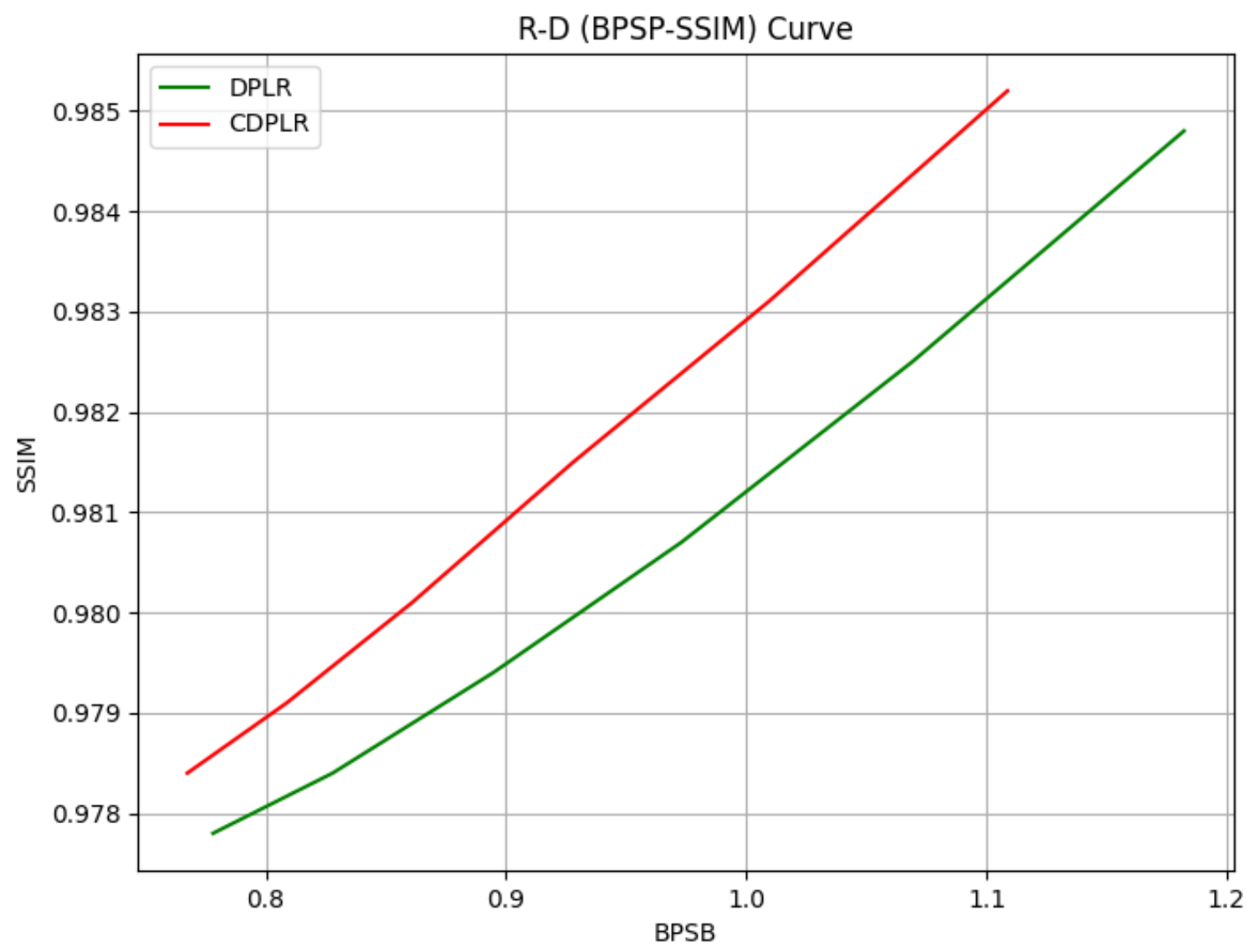

**Fig. 6.** R-D (BPSP-SSIM) curve of DLPR and CDLPR on single image of Kodak image dataset.

For the sake of exhibiting the effectiveness of the proposed CIC, the rate-distortion (R-D) curves of the DLPR and CDLPR algorithms on the single image, kodim01.png, of Kodak image dataset are shown in Figures 5 and 6. Figure 5 is the R-D (BPSP-PSNR) curve of DLPR and CDLPR and Figure 6 is the R-D (BPSP-SSIM) curve of DLPR and CDLPR. Figures 5 and 6 demonstrate that the proposed CDLPR significantly outperforms the classical DLPR in PSNR and SSIM at the same BPSP.

*4.2.2 Experimental Results of Reconstruction Performance*

This experiment is planned to compare the reconstruction performance between the traditional SIC and the proposed CIC. This experiment concentrates on the first five competing open-source methods in Table 3 and five public image datasets.

Figures 7, 8, and 9 demonstrate the PSNR, SSIM, and BPSP of DLPR and CDLPR algorithms with quantization parameter $\tau$=7 on Kodak image dataset. The experimental results indicate that the proposed CDLPR is superior to the classical DLPR in PSNR, SSIM, and BPSP.

Table 5 shows the PSNR, SSIM, and BPSP of DLPR and CDLPR with quantization parameter $\tau$=7 on five image datasets. The PSNR, SSIM, and BPSP are the averages on each dataset. $\Delta$ is the average increment of PSNR or SSIM between CDLPR and DLPR on each dataset. $\Delta_m$ is the maximum increment of PSNR or SSIM between CDLPR and DLPR on each dataset. $\Delta$ is also the average decrement of BPSP between CDLPR and DLPR on each dataset. $\Delta_m$ is also the maximum decrement of BPSP between CDLPR and DLPR on each dataset. $\sigma$ is the standard deviation of the PSNR or SSIM increment or the BPSP decrement. The maximum PSNR increment is 1.7142 dB on CLIC2021 Test image dataset. The experimental results manifest that the proposed CDLPR holds superiority over the classical DLPR in PSNR, SSIM, and BPSP.

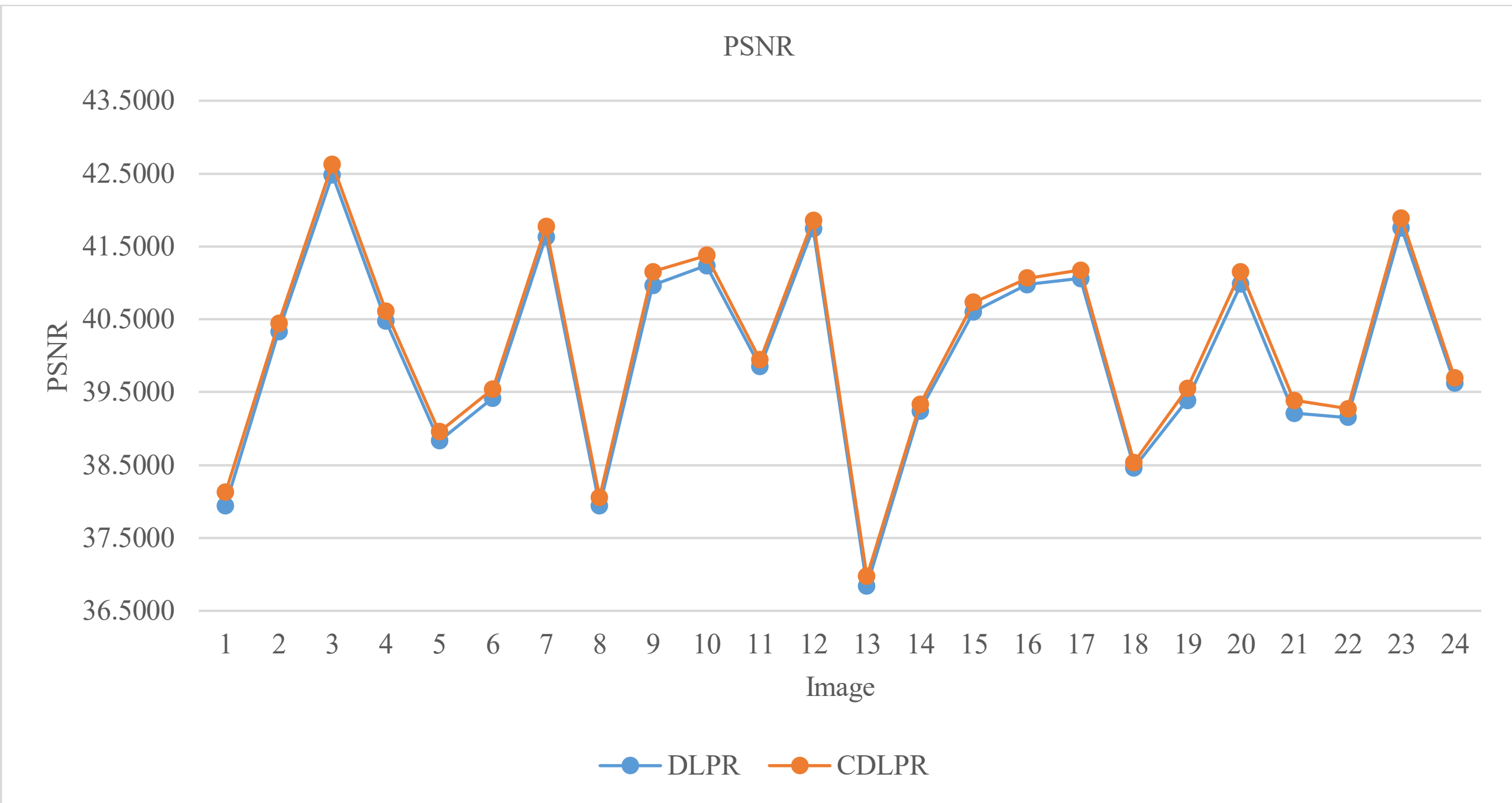

**Fig. 7.** PSNR of DLPR and CDLPR on Kodak image dataset.

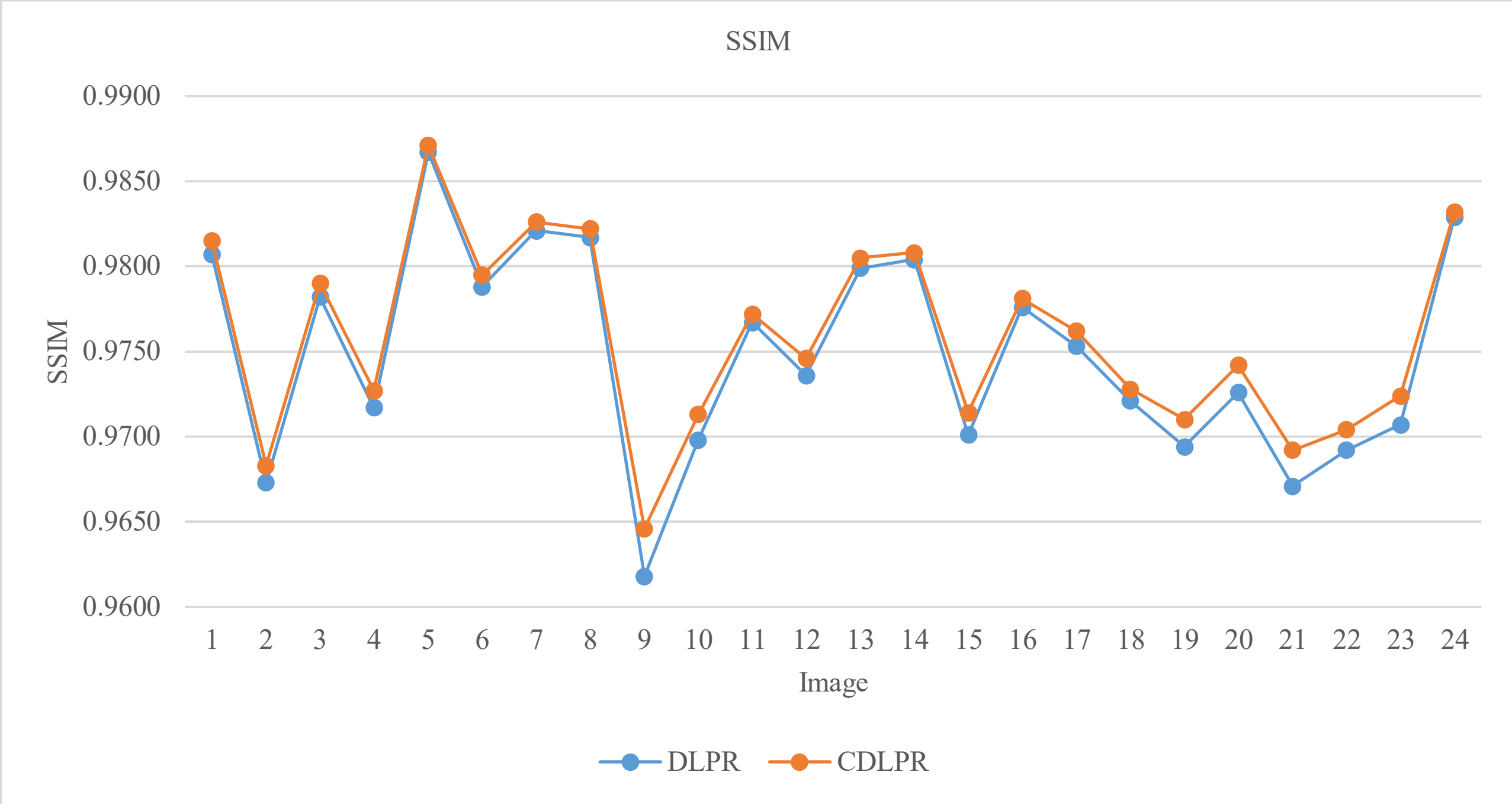

**Fig. 8.** SSIM of DLPR and CDLPR on Kodak image dataset.

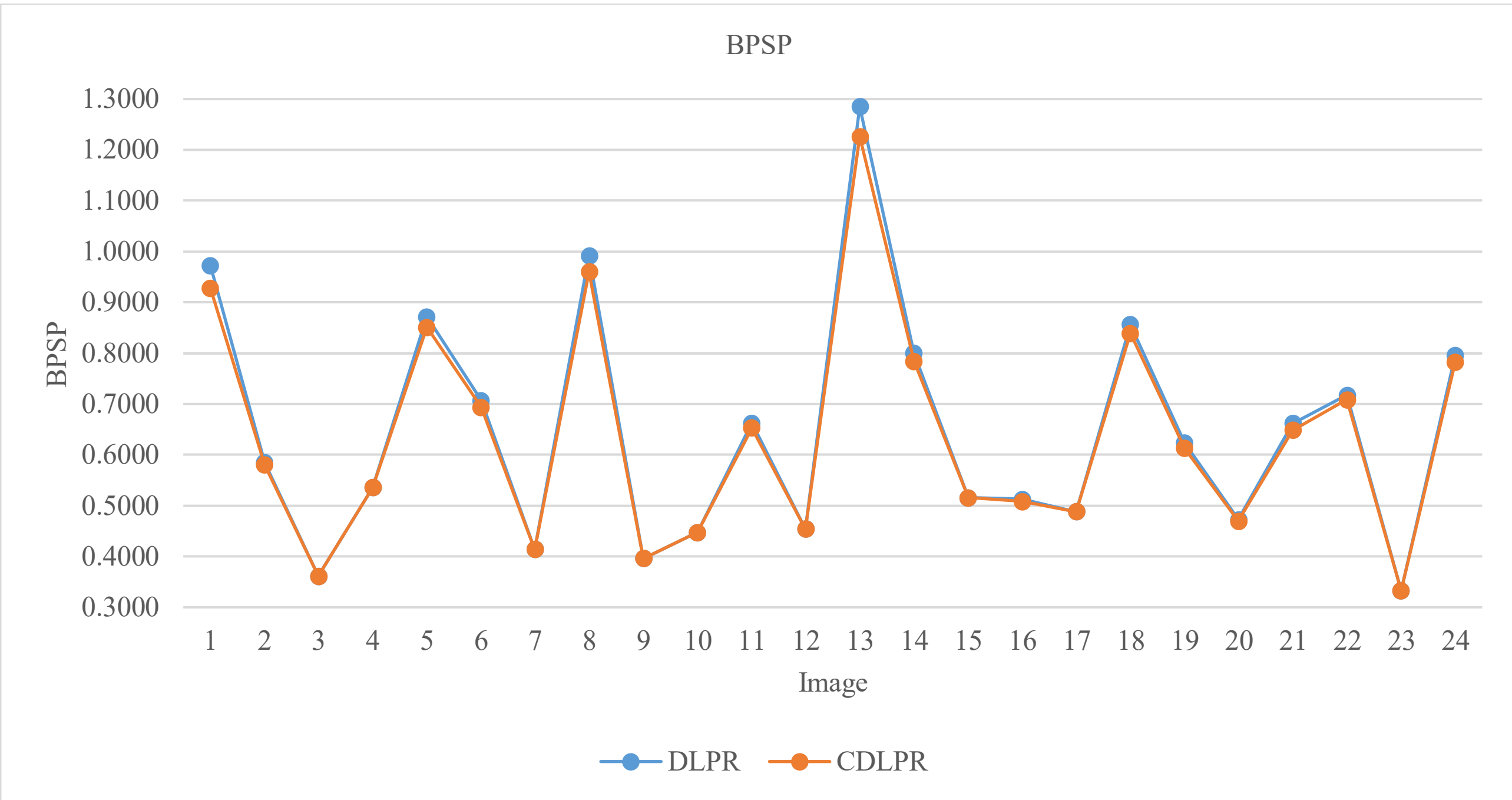

**Fig. 9.** BPSP of DLPR and CDLPR on Kodak image dataset.

Table 6 displays the PSNR, SSIM, and BPSP of CDC and CCDC with quantization parameter 0.0128 on five image datasets. The maximum PSNR increment is 5.7129 dB on CLIC2021 Test image dataset. For the convenience of computation, some images are clipped to the same size as images of Kodak image dataset. The experimental results reveal that the proposed CCDC overbalances the classical CDC in PSNR and SSIM while both of them have the same BPSR.

Table 5 EXPERIMENTAL RESULTS OF DLPR AND CDLPR ON FIVE IMAGE DATASETS

| Dataset | | Kodak | CLIC2021 Test | CLIC2021 Validation | CLIC2022 Validation | CLIC2024 Validation |
|---|---|---|---|---|---|---|
| PSNR↑ | DLPR | 40.0063 | 41.3359 | 40.5765 | 41.0578 | 40.2871 |
| | CDLPR | 40.1345 | 41.5588 | 40.7525 | 41.2347 | 40.4597 |
| | Δ | 0.1282 | 0.2229 | 0.1760 | 0.1769 | 0.1726 |
| | $\Delta_m$ | **0.1861** | **1.7142*** | **0.4736** | **0.3763** | **0.3537** |
| | σ | 0.0312 | 0.2068 | 0.0902 | 0.0725 | 0.0622 |
| SSIM↑ | DLPR | 0.9749 | 0.9711 | 0.9668 | 0.9714 | 0.9702 |
| | CDLPR | 0.9759 | 0.9739 | 0.9689 | 0.9728 | 0.9722 |
| | Δ | 0.0010 | 0.0027 | 0.0021 | 0.0014 | 0.0019 |
| | $\Delta_m$ | 0.0028 | 0.0388 | 0.0129 | 0.0054 | 0.0110 |
| | σ | 0.0006 | 0.0053 | 0.0027 | 0.0012 | 0.0024 |
| BPSP↓ | DLPR | 0.6443 | 0.4610 | 0.5089 | 0.5436 | 0.5920 |
| | CDLPR | 0.6329 | 0.4541 | 0.4999 | 0.5312 | 0.5802 |
| | Δ | 0.0114 | 0.0069 | 0.0090 | 0.0123 | 0.0119 |
| | $\Delta_m$ | 0.0585 | 0.0565 | 0.0577 | 0.0803 | 0.0522 |
| | σ | 0.0148 | 0.0124 | 0.0133 | 0.0182 | 0.0126 |

Table 6 EXPERIMENTAL RESULTS OF CDC AND CCDC ON FIVE IMAGE DATASETS

| Dataset | | Kodak | CLIC2021 Test | CLIC2021 Validation | CLIC2022 Validation | CLIC2024 Validation |
|---|---|---|---|---|---|---|
| PSNR↑ | CDC | 34.3532 | 38.2491 | 36.8646 | 37.0497 | 36.7745 |
| | CCDC | 34.4227 | 38.7852 | 37.0708 | 37.6947 | 37.0449 |
| | Δ | 0.0694 | 0.5361 | 0.2062 | 0.6450 | 0.2704 |
| | $\Delta_m$ | **0.1448** | **5.7129*** | **1.8855** | **2.8527** | **2.8118** |
| | σ | 0.0447 | 1.0812 | 0.3339 | 0.8858 | 0.5096 |
| SSIM↑ | CDC | 0.9379 | 0.9447 | 0.9364 | 0.9409 | 0.9387 |
| | CCDC | 0.9382 | 0.9466 | 0.9373 | 0.9417 | 0.9405 |
| | Δ | 0.0003 | 0.0019 | 0.0009 | 0.0008 | 0.0018 |
| | $\Delta_m$ | 0.0015 | 0.0503 | 0.0144 | 0.0095 | 0.0131 |
| | σ | 0.0003 | 0.0071 | 0.0026 | 0.0017 | 0.0033 |
| BPSP↓ | CDC | 0.8389 | 0.3926 | 0.5368 | 0.4751 | 0.4815 |
| | CCDC | 0.8389 | 0.3926 | 0.5368 | 0.4751 | 0.4815 |
| | Δ | 0.0000 | 0.0000 | 0.0000 | 0.0000 | 0.0000 |
| | $\Delta_m$ | 0.0000 | 0.0000 | 0.0000 | 0.0000 | 0.0000 |
| | σ | 0.0000 | 0.0000 | 0.0000 | 0.0000 | 0.0000 |

Table 7 EXPERIMENTAL RESULTS OF ICT AND CICT ON FIVE IMAGE DATASETS

| Dataset | | Kodak | CLIC2021 Test | CLIC2021 Validation | CLIC2022 Validation | CLIC2024 Validation |
|---|---|---|---|---|---|---|
| PSNR↑ | ICT | 29.4609 | 29.1243 | 29.6862 | 29.3554 | 28.1825 |
| | CICT | 29.5614 | 29.2873 | 29.7799 | 29.4643 | 28.2794 |
| | Δ | 0.1005 | 0.1630 | 0.0936 | 0.1089 | 0.0969 |
| | $\Delta_m$ | **0.2573** | **2.7907*** | **0.2281** | **0.2830** | **0.2254** |
| | σ | 0.0489 | 0.3895 | 0.0507 | 0.0655 | 0.0538 |
| SSIM↑ | ICT | 0.7191 | 0.7994 | 0.7812 | 0.7940 | 0.7931 |
| | CICT | 0.7254 | 0.8042 | 0.7856 | 0.7984 | 0.7979 |
| | Δ | 0.0062 | 0.0048 | 0.0045 | 0.0045 | 0.0048 |
| | $\Delta_m$ | 0.0112 | 0.0237 | 0.0107 | 0.0142 | 0.0305 |
| | σ | 0.0026 | 0.0038 | 0.0022 | 0.0033 | 0.0053 |
| BPSP↓ | ICT | 0.4007 | 0.3947 | 0.3445 | 0.4665 | 0.5397 |
| | CICT | 0.3313 | 0.3891 | 0.3395 | 0.4058 | 0.4683 |
| | Δ | 0.0694 | 0.0056 | 0.0049 | 0.0607 | 0.0715 |
| | $\Delta_m$ | 0.1378 | 0.0976 | 0.1155 | 0.1097 | 0.1148 |
| | σ | 0.0235 | 0.0200 | 0.0198 | 0.0269 | 0.0231 |

Table 8 EXPERIMENTAL RESULTS OF QRVAE AND CQRVAE ON FIVE IMAGE DATASETS

| Dataset | | Kodak | CLIC2021 Test | CLIC2021 Validation | CLIC2022 Validation | CLIC2024 Validation |
|---|---|---|---|---|---|---|
| PSNR↑ | QRVAE | 30.0170 | 34.0337 | 32.3537 | 33.4110 | 32.6968 |
| | CQRVAE | 30.9856 | 35.2369 | 33.3826 | 34.4798 | 33.8661 |
| | Δ | 0.9686 | 1.2032 | 1.0289 | 1.0688 | 1.1693 |
| | $\Delta_m$ | **1.2953** | **4.9347*** | **2.5864** | **4.3077** | **3.6048** |
| | σ | 0.1317 | 0.8415 | 0.4206 | 0.8423 | 0.7411 |
| SSIM↑ | QRVAE | 0.8093 | 0.8937 | 0.8625 | 0.8856 | 0.8873 |
| | CQRVAE | 0.8599 | 0.9102 | 0.8887 | 0.9013 | 0.9050 |
| | Δ | 0.0506 | 0.0164 | 0.0262 | 0.0157 | 0.0177 |
| | $\Delta_m$ | 0.0966 | 0.0991 | 0.1094 | 0.1028 | 0.0533 |
| | σ | 0.0199 | 0.0189 | 0.0259 | 0.0208 | 0.0150 |
| BPSP↓ | QRVAE | 0.1829 | 0.0653 | 0.1008 | 0.0948 | 0.0860 |
| | CQRVAE | 0.1829 | 0.0653 | 0.1008 | 0.0948 | 0.0860 |
| | Δ | 0.0000 | 0.0000 | 0.0000 | 0.0000 | 0.0000 |
| | $\Delta_m$ | 0.0000 | 0.0000 | 0.0000 | 0.0000 | 0.0000 |
| | σ | 0.0000 | 0.0000 | 0.0000 | 0.0000 | 0.0000 |

Table 9 EXPERIMENTAL RESULTS OF VQVAE AND CVQVAE ON FIVE IMAGE DATASETS

| Dataset | | Kodak | CLIC2021 Test | CLIC2021 Validation | CLIC2022 Validation | CLIC2024 Validation |
|---|---|---|---|---|---|---|
| PSNR↑ | VQVAE | 32.1650 | 32.5532 | 32.8301 | 32.6257 | 33.1952 |
| | CVQVAE | 32.5840 | 32.9286 | 33.2335 | 32.9414 | 33.5463 |
| | Δ | 0.4190 | 0.3755 | 0.4034 | 0.3157 | 0.3511 |
| | $\Delta_m$ | **1.1633** | **0.8489** | **1.5219*** | **1.2535** | **0.8269** |
| | σ | 0.2132 | 0.2037 | 0.2923 | 0.2483 | 0.2214 |
| SSIM↑ | VQVAE | 0.9657 | 0.9613 | 0.9578 | 0.9685 | 0.9576 |
| | CVQVAE | 0.9663 | 0.9626 | 0.9586 | 0.9690 | 0.9601 |
| | Δ | 0.0006 | 0.0013 | 0.0009 | 0.0005 | 0.0025 |
| | $\Delta_m$ | 0.0020 | 0.0195 | 0.0082 | 0.0021 | 0.0397 |
| | σ | 0.0005 | 0.0031 | 0.0016 | 0.0006 | 0.0071 |
| BPSP↓ | VQVAE | 0.9936 | 0.6831 | 0.7546 | 0.7986 | 0.8077 |
| | CVQVAE | 0.9936 | 0.6831 | 0.7546 | 0.7986 | 0.8077 |
| | Δ | 0.0000 | 0.0000 | 0.0000 | 0.0000 | 0.0000 |
| | $\Delta_m$ | 0.0000 | 0.0000 | 0.0000 | 0.0000 | 0.0000 |
| | σ | 0.0000 | 0.0000 | 0.0000 | 0.0000 | 0.0000 |

Table 7 illustrates the PSNR, SSIM, and BPSP of ICT and CICT with quantization parameter 3 on five image datasets. The maximum PSNR increment is 2.7907 dB on CLIC2021 Test image dataset. For the expediency of calculation, some images are trimmed to the same size as images of Kodak image dataset. The experimental results uncover that the proposed CICT overmatches the classical ICT in PSNR, SSIM, and BPSP.

Table 8 demonstrates the PSNR, SSIM, and BPSP of QRVAE and CQRVAE with quantization parameter 16 on five image datasets. The maximum PSNR increment is 4.9347 dB on CLIC2021 Test image dataset. For the facilitation of implementation, some images are tailored to the same size of images of Kodak image dataset. The experimental results indicate that the proposed CQRVAE outperforms the classical QRVAE in PSNR and SSIM while both of them have the same BPSR.

Table 9 exhibits the PSNR, SSIM, and BPSP of VQVAE and CVQVAE on five image datasets. The maximum PSNR increment is 1.5219 dB on CLIC2021 Validation image dataset. For the easiness of realization, some images are cropped to the same size as images of Kodak image dataset. The experimental results make clear that the proposed CVQVAE surpasses the classical VQVAE in PSNR and SSIM while both of them have the same BPSR.

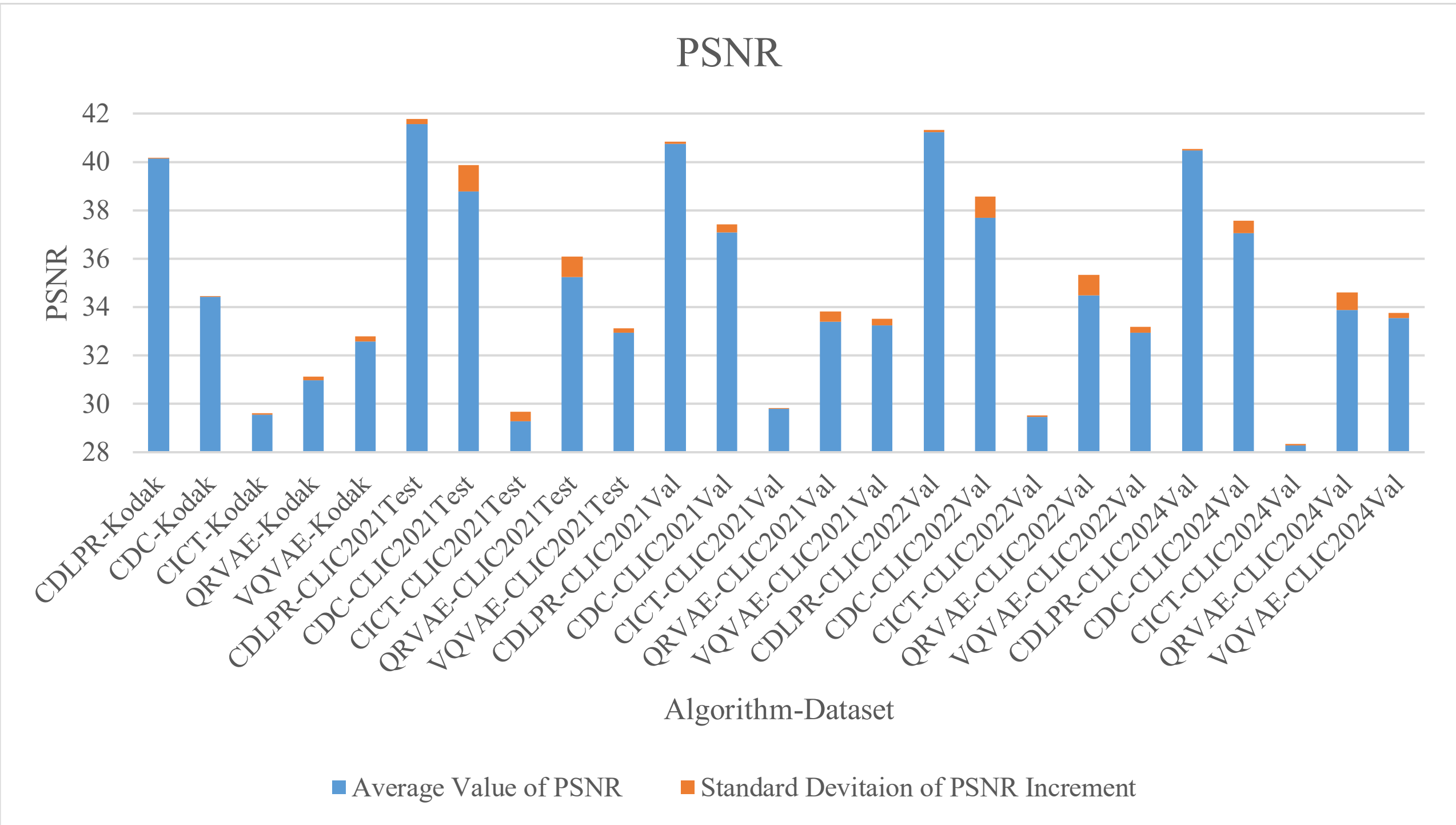

**Fig. 10.** Average value of PSNR and standard deviation of PSNR increment for the first five competing algorithms on five datasets.

For the purpose of vividly presenting PSNR increment, Fig. 10 shows the average value of PSNR and the standard deviation of PSNR increment for the first five competing algorithms on five datasets. The average value and standard deviation determine confidence intervals.

In order to explicitly show the performance difference of image reconstruction between the proposed CIC and the classical SIC, some example images are displayed in Figures 11 to 15.

Figure 11 shows the experimental results of DLPR and CDLPR with maximum PSNR increment on Kodak image dataset. Figure 11(a) is the original image, Figure 11(b) is the reconstruction image of DLPR, and Figure 11(c) is the reconstruction image of CDLPR. It is hard to discover the difference between Figure 11(b) and Figure 11(c). Figure 11(d) is the subblock of Figure 11(a), Figure 11(e) is the related subblock of Figure 11(b), and Figure 11(f) is the related subblock of Figure 11(c). Figure 11(d), (e), and (f) are marked with red boxes in Figure 11(a), (b), and (c) respectively. It is also hard to discover the difference between Figure 11(e) and Figure 11(f). Figure 11(g) is the absolute difference image block between Figure 11(e) and Figure 11(d), and Figure 11(h) is the absolute difference image block between Figure 11(f) and Figure 11(d). It is still hard to discover the difference between Figure 11(g) and Figure 11(h). Figure 11(i) is the logic difference image block between Figure 11(e) and Figure 11(d), and Figure 11(j) is the logic difference image block between Figure 11(f) and Figure 11(d). It is easy to discover the difference between Figure 11(i) and Figure 11(j). Figure 11(i) and Figure 11(j) indicate that the reconstruction image quality of CDLPR outperforms that of DLPR. Figure 11 indicates that the proposed method is effective for testing images with sharp edges.

Figure 12 displays the experimental results of DLPR and CDLPR with maximum PSNR rise on CLIC2021 test image dataset. The layout of Figure 12 is same as Figure 11. It is difficult to seek out the discrepancy between Figure 12(b) and Figure 12(c); It is also difficult to seek out the discrepancy between Figure 12(e) and Figure 12(f); It is still difficult to seek out the discrepancy between Figure 12(g) and Figure 12(h); It is effortless to seek out the discrepancy between Figure 12(i) and Figure 12(j). Figure 12(i) and Figure 12(j) reveal that the restoration image quality of CDLPR outbalances that of DLPR. Figure 12 shows that the proposed method is propitious to testing images with dark backgrounds and high contrast.

Figure 13 illustrates the experimental results of DLPR and CDLPR with maximum PSNR increase on CLIC2021 validation image dataset. The composition of Figure 13 is same as Figure 11. It is tough to check the discrimination between Figure 13(b) and Figure 13(c); It is also tough to check the discrimination between Figure 13(e) and Figure 13(f); It is still tough to check the discrimination between Figure 13(g) and Figure 13(h); It is toil-less to check the discrimination between Figure 13(i) and Figure 13(j). Figure 13(i) and Figure 13(j) uncover that the recovery image quality of CDLPR overmatches that of DLPR. Figure 13 demonstrates that the proposed method is appropriate for testing images with grid shapes.

Figure 14 demonstrates the experimental results of DLPR and CDLPR with maximum PSNR improvement on CLIC2022 validation image dataset. The arrangement of Figure 14 is same as

Figure 11. It is arduous to examine the distinction between Figure 14(b) and Figure 14(c); It is also arduous to examine the distinction between Figure 14(e) and Figure 14(f); It is still arduous to examine the distinction between Figure 14(g) and Figure 14(h); It is convenient to examine the distinction between Figure 14(i) and Figure 14(j). Figure 14(i) and Figure 12(j) disclose that the reestablishment image quality of CDLPR exceeds that of DLPR. Figure 14 indicates that the proposed method is fit for testing images with dark backgrounds and high contrast.

Figure 15 exhibits the experimental results of DLPR and CDLPR with maximum PSNR gain on CLIC2024 validation image dataset. The organization of Figure 15 is same as Figure 11. It is formidable to inspect the distinguishing between Figure 15(b) and Figure 15(c); It is also formidable to inspect the distinguishing between Figure 15(e) and Figure 15(f); It is still formidable to inspect the distinguishing between Figure 15(g) and Figure 15(h); It is facile to inspect the distinguishing between Figure 15(i) and Figure 15(j). Figure 15(i) and Figure 15(j) expose that the rebuilding image quality of CDLPR surpasses that of DLPR. Figure 15 shows the proposed method is suitable for testing images with complicated patterns.

Therefore, the proposed CIC holds superiority over the classical SIC in reconstruction performance and is especially appropriate for testing images with sharp edges, dark backgrounds, high contrast, grid shapes, and complicated patterns.

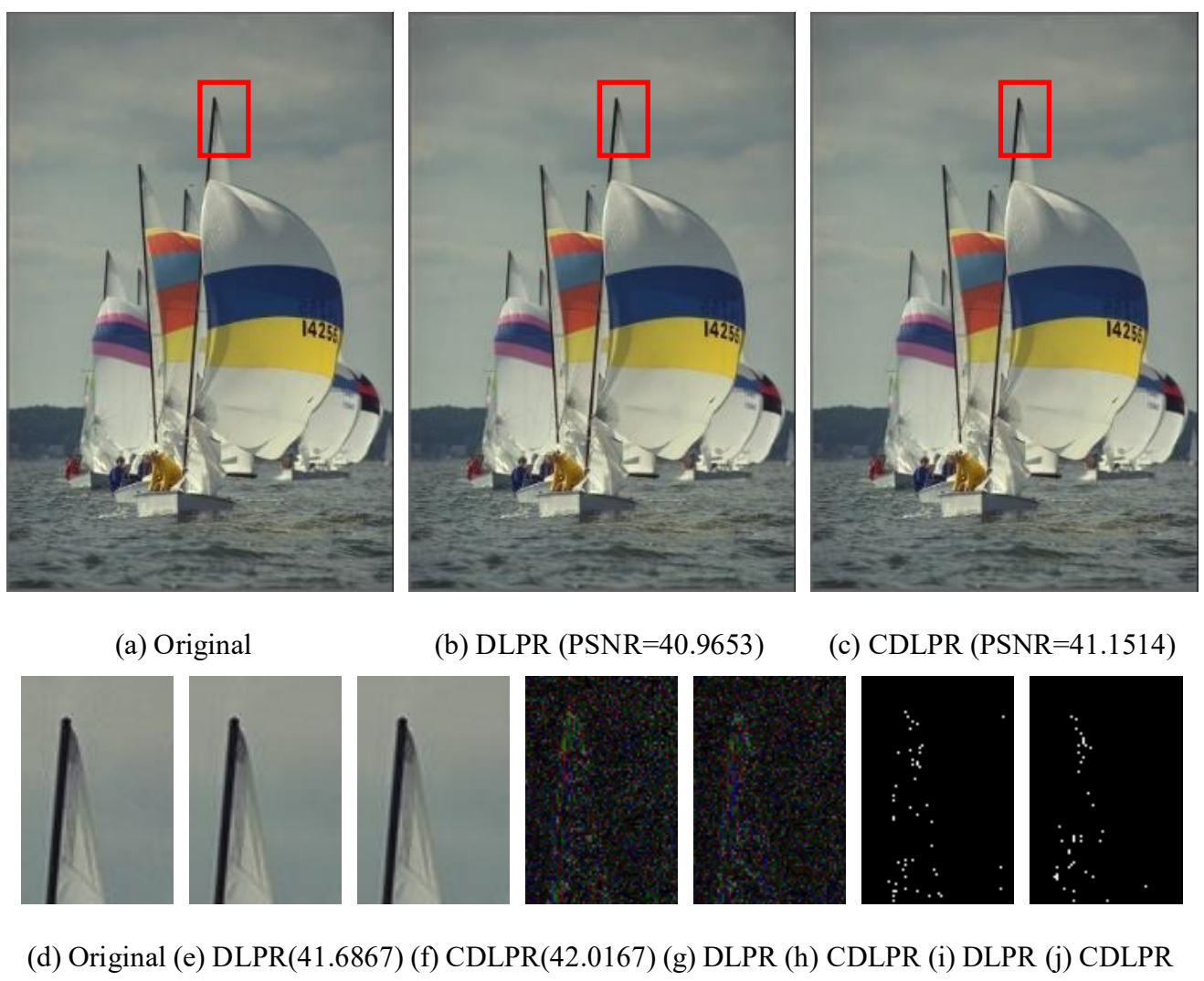

**Fig. 11.** Experimental results of DLPR and CDLPR with maximum PSNR increment on Kodak image with sharp edges.

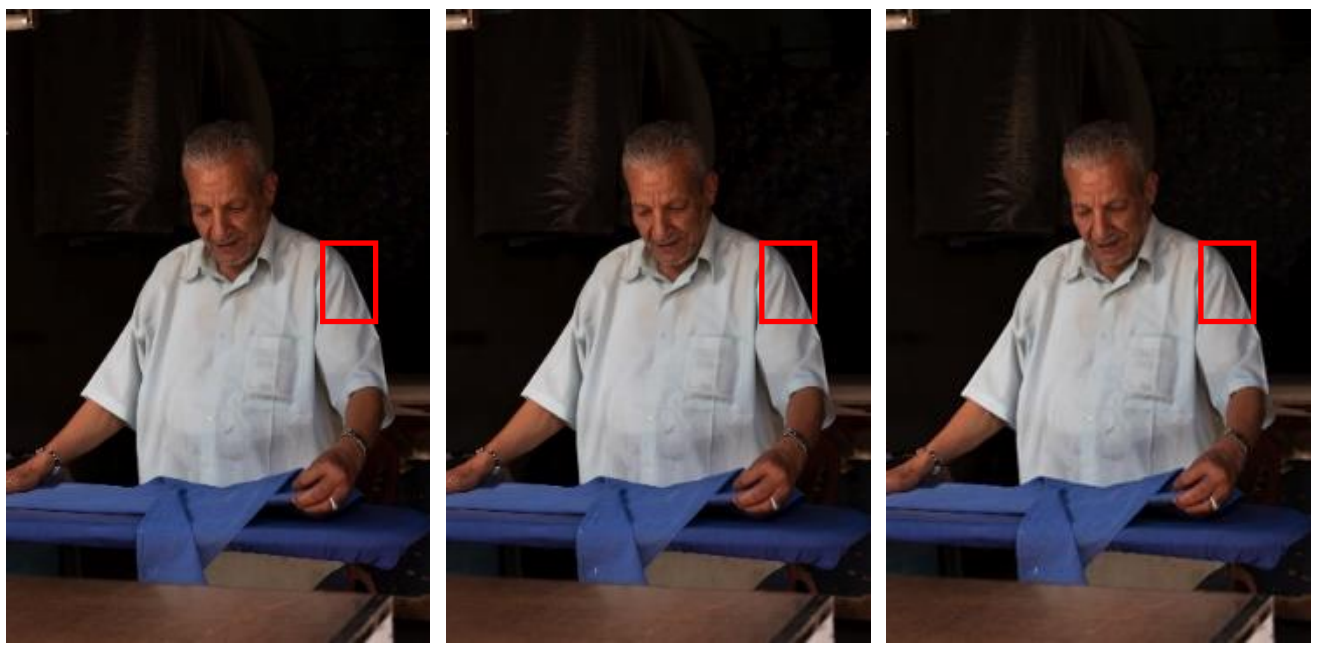

(a) Original (b) DLPR (PSNR=43.2816) (c) CDLPR (PSNR=44.9958)

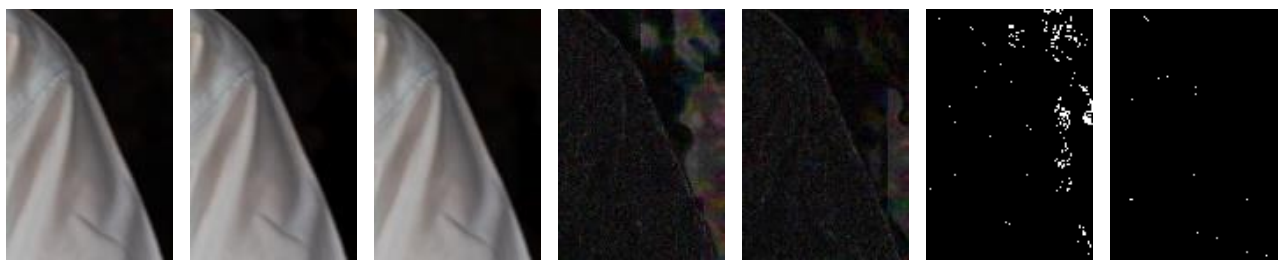

(d) Original (e) DLPR(41.5044) (f) CDLPR(43.5667) (g) DLPR (h) CDLPR (i) DLPR (j) CDLPR

**Fig. 12.** Experimental results of DLPR and CDLPR with maximum PSNR increment on CLIC2021 test image with dark backgrounds and high contrast.

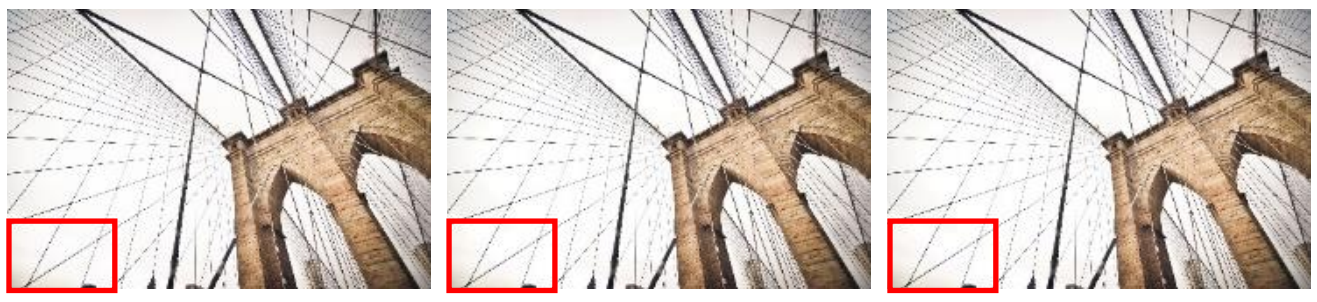

(a) Original (b) DLPR (PSNR=39.9442) (c) CDLPR (PSNR=40.4178)

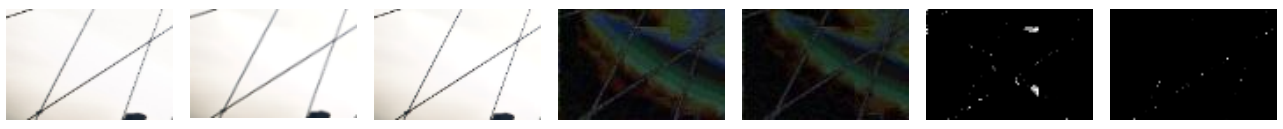

(d) Original (e) DLPR(40.6533) (f) CDLPR(42.0502) (g) DLPR (h) CDLPR (i) DLPR (j) CDLPR

**Fig. 13.** Experimental results of DLPR and CDLPR with maximum PSNR increment on CLIC2021 validation image with grid shapes.

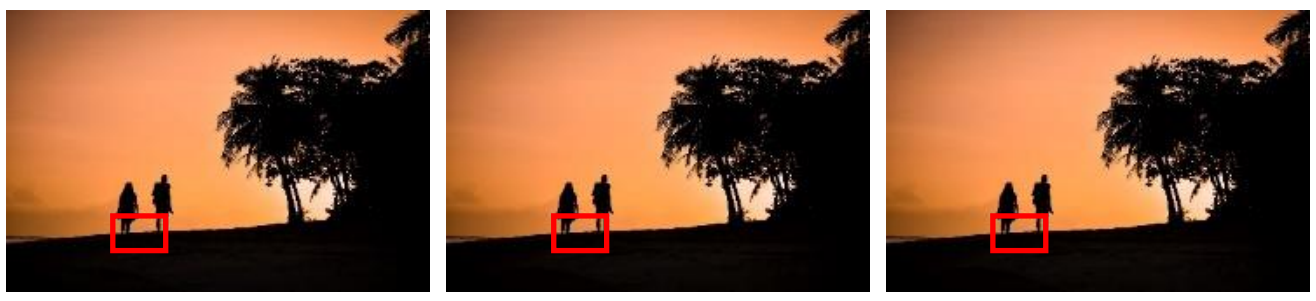

(a) Original (b) DLPR (PSNR=45.2744) (c) CDLPR (PSNR=45.6507)

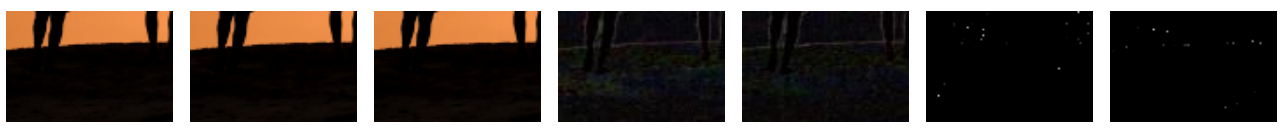

(d) Original (e) DLPR(44.7260) (f) CDLPR(45.4461) (g) DLPR (h) CDLPR (i) DLPR (j) CDLPR

**Fig. 14.** Experimental results of DLPR and CDLPR with maximum PSNR increment on CLIC2022 validation image with dark backgrounds and high contrast.

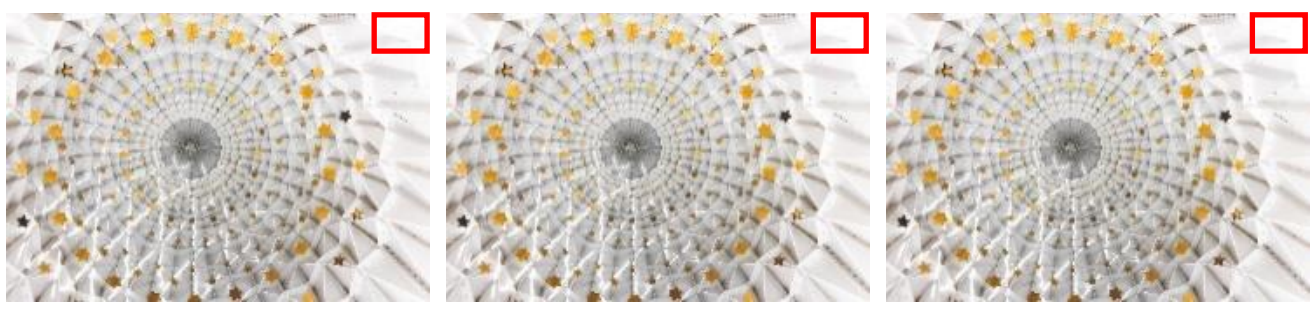

(a) Original (b) DLPR (PSNR=43.1443) (c) CDLPR (PSNR=43.4980)

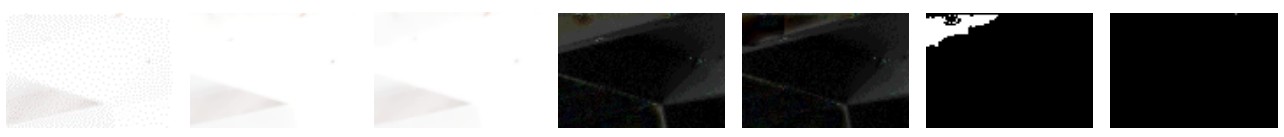

(d) Original (e) DLPR(42.8405) (f) CDLPR(44.6408) (g) DLPR (h) CDLPR (i) DLPR (j) CDLPR

**Fig. 15.** Experimental results of DLPR and CDLPR with maximum PSNR increment on CLIC2024 validation image with complicated patterns.

*4.2.3 Experiment Results of Out-of-Sample*

This experiment is arranged to assess the reconstruction performance for out-of-sample testing images. Five typical out-of-sample testing images are selected in Figures 16 to 20. DLPR and CDLPR algorithms with quantization parameter $\tau$=7 are chosen for performance comparison. The experimental results are listed in Table 10 and shown in Figures 16 to 20. $\Delta_s$ is the maximum increment of subblock PSNR or SSIM.

Figure 16 offers the experimental results of out-of-sample testing image 1. Figure 16(a) is the original image, Figure 16(b) is the reconstruction image of DLPR, and Figure 16(c) is the reconstruction image of CDLPR. It is untoward to survey the difference between Figure 16(b) and Figure 16(c). Figure 16(d) is the subblock of Figure 16(a), Figure 16(e) is the related subblock of Figure 16(b), and Figure 16(f) is the related subblock of Figure 16(c). Figure 16(d), (e), and (f) are marked with red boxes in Figure 16(a), (b), and (c) respectively. It is also untoward to survey the difference between Figure 16(e) and Figure 16(f). Figure 16(g) is the absolute difference image block between Figure 16(e) and Figure 16(d), and Figure 16(h) is the absolute difference image block between Figure 16(f) and Figure 16(d). It is still untoward to survey the difference between Figure 16(g) and Figure 16(h). Figure 16(i) is the logic difference image block between Figure 16(e) and Figure 16(d), and Figure 16(j) is the logic difference image block between Figure 16(f) and Figure 16(d). It is undemanding to survey the difference between Figure 16(i) and Figure 16(j). Figure 16(i) and Figure 16(j) unmask that the reconstruction image quality of CDLPR outperforms that of DLPR. Figure 16 unmasks that the proposed method is effective for out-of-sample testing images with dark backgrounds and detailed foregrounds.

Table 10 EXPERIMENTAL RESULTS OF DLPR AND CDLPR FOR OUT-OF-SAMPLE TESTING IMAGES

| Testing Images | | 1 | 2 | 3 | 4 | 5 |
|---|---|---|---|---|---|---|
| PSNR↑ | DLPR | 42.0104 | 42.0685 | 41.5296 | 40.7594 | 42.7895 |
| | CDLPR | 42.9432 | 42.7137 | 41.9656 | 41.7546 | 43.3530 |
| | Δ | 0.9329 | 0.6452 | 0.4361 | 0.9953 | 0.5635 |
| | $\Delta_s$ | **1.1742** | **0.9036** | **0.7093** | **2.4978*** | **1.0996** |
| SSIM↑ | DLPR | 0.9818 | 0.9308 | 0.8477 | 0.8731 | 0.9850 |
| | CDLPR | 0.9885 | 0.9460 | 0.8599 | 0.9088 | 0.9888 |
| | Δ | 0.0067 | 0.0152 | 0.0122 | 0.0357 | 0.0039 |
| | $\Delta_s$ | 0.0111 | 0.0421 | 0.0221 | 0.0314 | 0.0068 |
| BPSP↓ | DLPR | 0.4892 | 0.5833 | 0.2888 | 0.5056 | 0.6678 |
| | CDLPR | 0.4840 | 0.5692 | 0.2851 | 0.4986 | 0.6633 |
| | Δ | 0.0052 | 0.0141 | 0.0037 | 0.0070 | 0.0045 |

Figure 17 provides the experimental results of out-of-sample testing image 2. Figure 17 holds identical layout as Figure 16. It is stiff to inquiry the discrepancy between Figure 17(b) and Figure 17(c); It is also stiff to inquiry the discrepancy between Figure 17(e) and Figure 17(f); It is still stiff to inquiry the discrepancy between Figure 17(g) and Figure 17(h); It is simple to inquiry the discrepancy between Figure 17(i) and Figure 17(j). Figure 17(i) and Figure 17(j) state that the restoration image quality of CDLPR outbalances that of DLPR. Figure 17 states that the proposed method is propitious to out-of-sample testing images with dark backgrounds and delicate foregrounds.

Figure 18 affords the experimental results of out-of-sample testing image 3. Figure 18 holds identical composition as Figure 16. It is burdensome to investigate the discrimination between Figure 18(b) and Figure 18(c); It is also burdensome to investigate the discrimination between Figure 18(e) and Figure 18(f); It is still burdensome to investigate the discrimination between Figure 18(g) and Figure 18(h); It is straightforward to investigate the discrimination between Figure 18(i) and Figure 18(j). Figure 18(i) and Figure 18(j) declare that the recovery image quality of CDLPR overmatches that of DLPR. Figure 18 declares that the proposed method is appropriate for out-of-sample testing images with dark backgrounds and exquisite foregrounds.

Figure 19 furnishes the experimental results of out-of-sample testing image 4. Figure 19 holds identical arrangement as Figure 16. It is rough to research the distinction between Figure 19(b) and Figure 19(c); It is also rough to research the distinction between Figure 19(e) and Figure 17(f); It is still rough to research the distinction between Figure 19(g) and Figure 19(h); It is explicit to research the distinction between Figure 19(i) and Figure 19(j). Figure 19(i) and Figure 19(j) express that the reestablishment image quality of CDLPR exceeds that of DLPR. Figure 19 expresses that the proposed method is fit for out-of-sample testing images with dark backgrounds and intricate patterns.

Figure 20 supplies the experimental results of out-of-sample testing image 5. Figure 20 holds identical organization as Figure 16. It is onerous to look up the distinguishing between Figure 20(b) and Figure 20(c); It is also onerous to look up the distinguishing between Figure 20(e) and Figure 20(f); It is still onerous to look up the distinguishing between Figure 20(g) and Figure 20(h); It is unambiguous to look up the distinguishing between Figure 20(i) and Figure 20(j). Figure 20(i) and Figure 20(j) present that the rebuilding image quality of CDLPR surpasses that of DLPR. Figure 20 presents the proposed method is suitable for out-of-sample testing images with high contrast and tangled patterns.

*4.2.4 Experiment Results of the Latest Competing Algorithms*

This experiment is intended to compare the reconstruction performance between the latest SIC and the proposed CIC. This experiment focuses on the last three competing open-source algorithms in Table 3 and Kodak dataset.

Table 11 shows the average PSNR, SSIM, and BPSP of CGIC and CCGIC with quantization parameter 0.4. “Original” means the classical SIC and “Circular” means the related CIC. The maximum PSNR increment is 0.5037 dB. Table 11 also shows the average PSNR, SSIM, and BPSP of TACO and CTACO with quantization parameter λ=0.0004. The maximum PSNR increment is 0.1806 dB. Table 11 further shows the average PSNR, SSIM, and BPSP of LBIC and CLBIC with quantization parameter 117.045. The maximum PSNR increment is 11.7911 dB. The experimental results demonstrate that the proposed CCGIC, CTACO, and CLBIC are superior to the traditional CGIC, TACO, and LBIC in PSNR, SSIM, and BPSP.

All in all, the proposed CIC is superior to the classical SIC in reconstruction performance and is especially suitable for out-of-sample testing images with dark backgrounds, detailed foregrounds, complicated patterns, and high contrast.

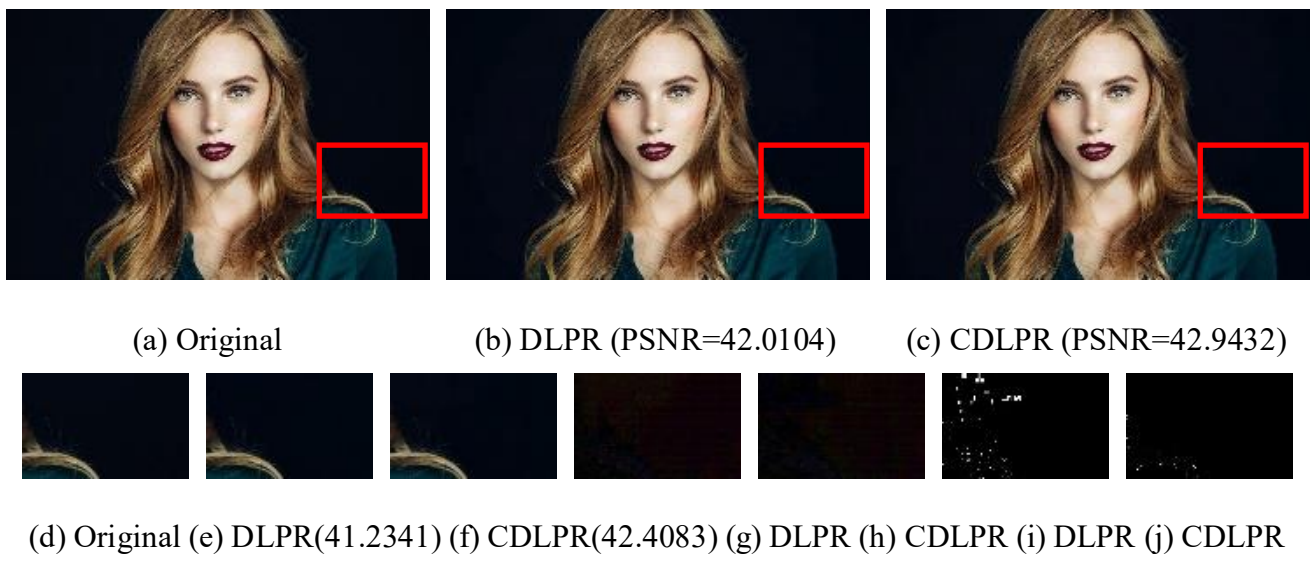

**Fig. 16.** Experimental results of DLPR and CDLPR for out-of-sample testing image 1 with dark backgrounds and detailed foregrounds.

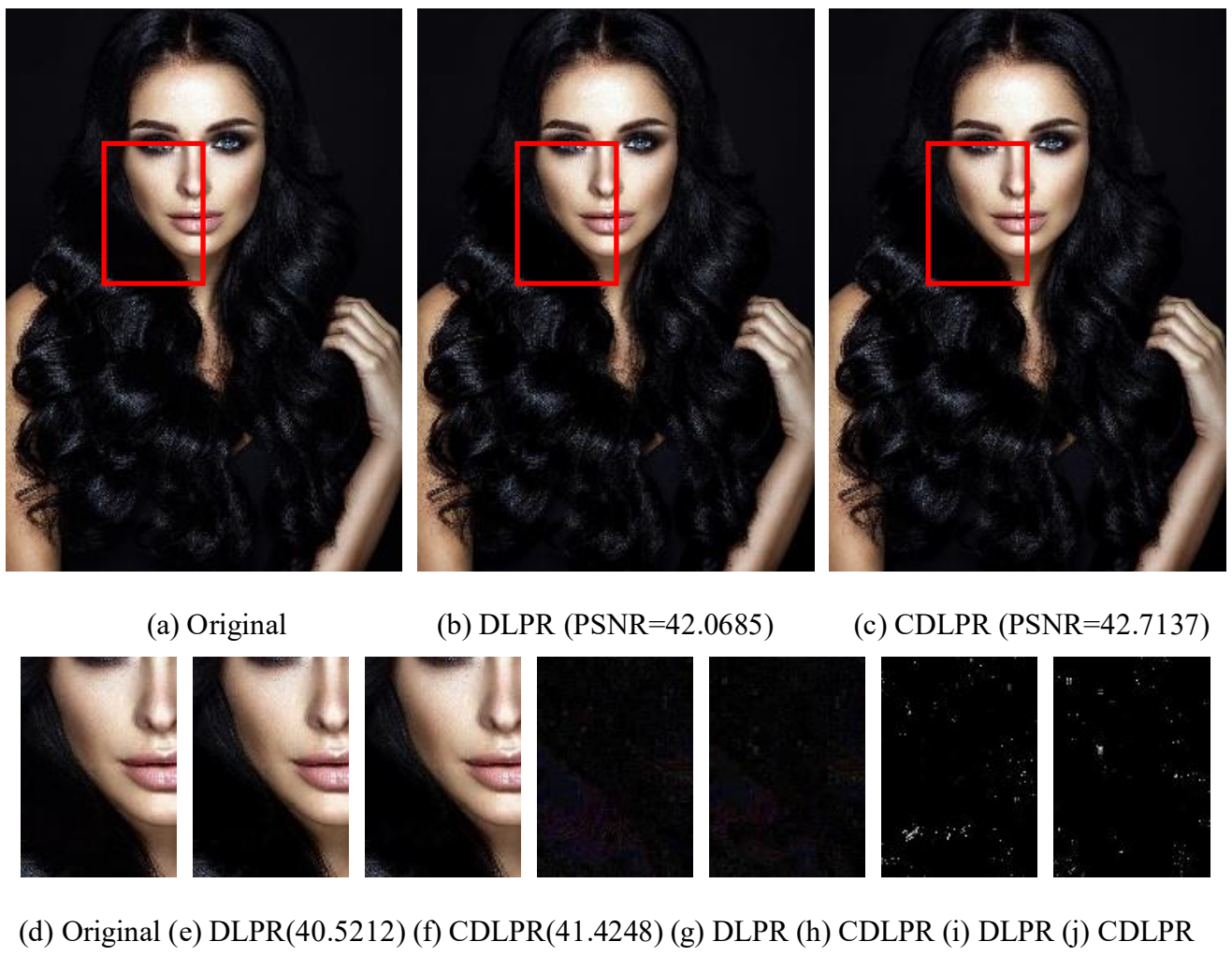

**Fig. 17.** Experimental results of DLPR and CDLPR for out-of-sample testing image 2 with dark backgrounds and delicate foregrounds.

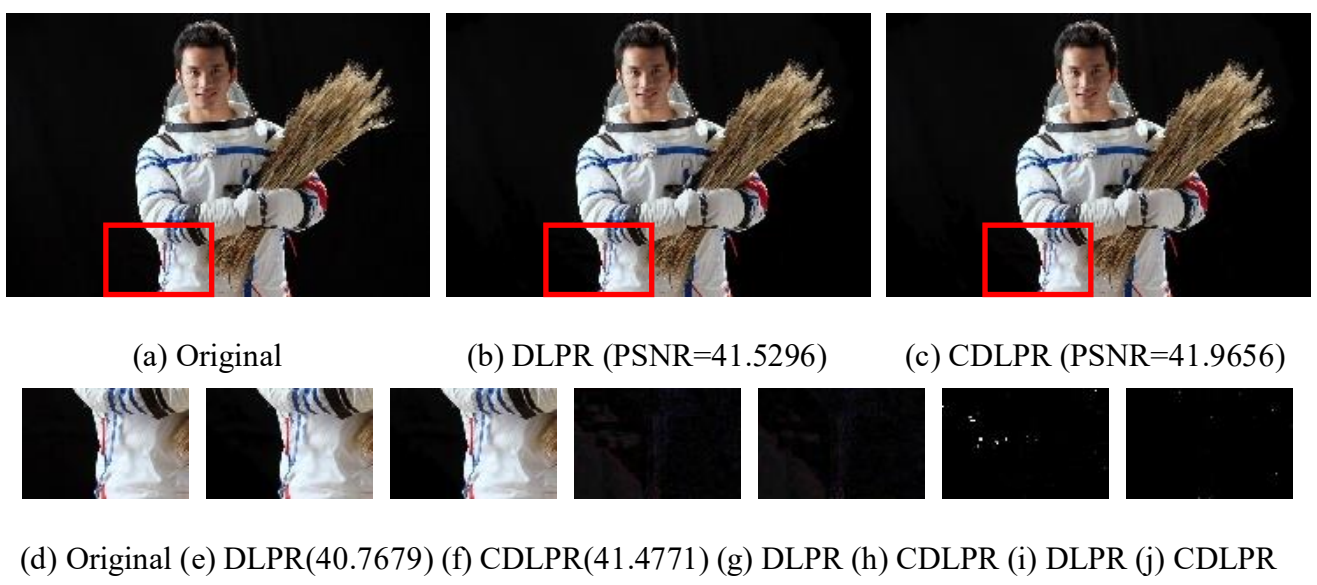

**Fig. 18.** Experimental results of DLPR and CDLPR for out-of-sample testing image 3 with dark backgrounds and exquisite foregrounds.

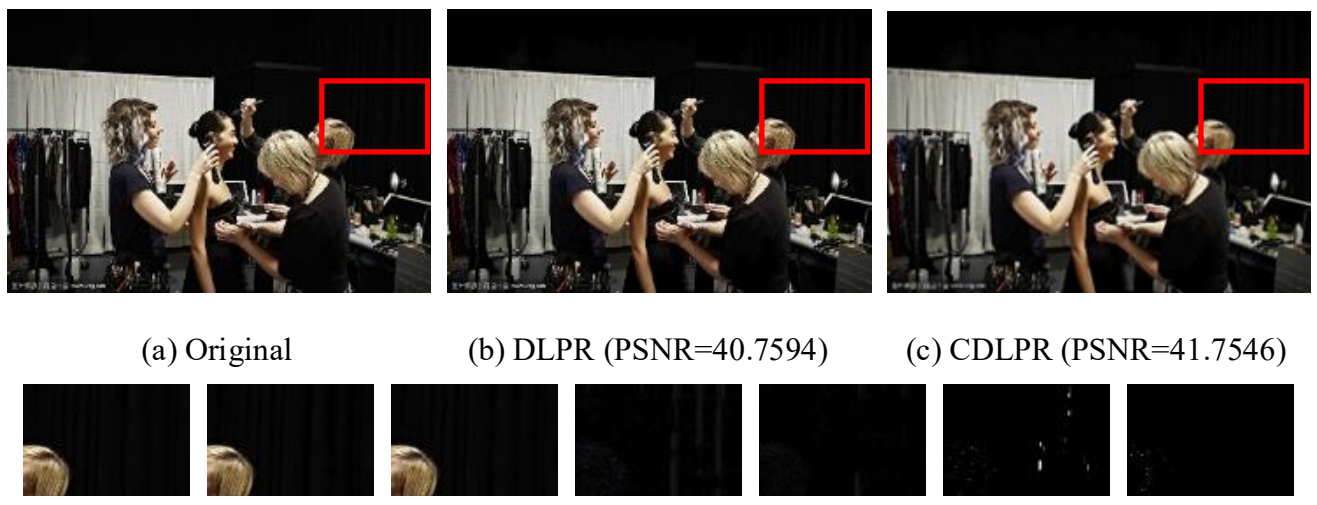

(a) Original (b) DLPR (PSNR=40.7594) (c) CDLPR (PSNR=41.7546)

(d) Original (e) DLPR(44.7458) (f) CDLPR(47.2436) (g) DLPR (h) CDLPR (i) DLPR (j) CDLPR

**Fig. 19.** Experimental results of DLPR and CDLPR for out-of-sample testing image 4 with dark backgrounds and intricate patterns.

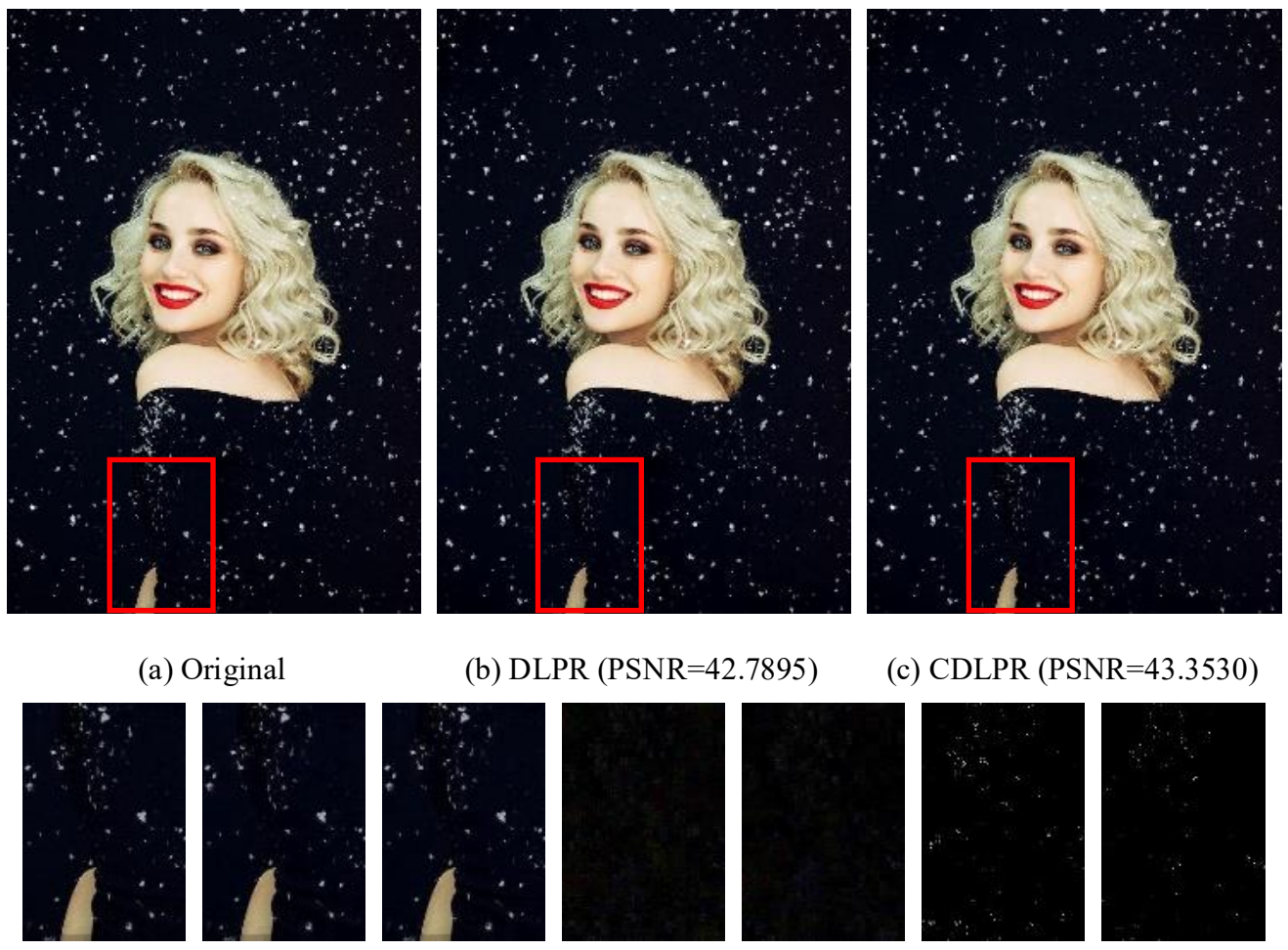

(a) Original (b) DLPR (PSNR=42.7895) (c) CDLPR (PSNR=43.3530)

(d) Original (e) DLPR(42.5610) (f) CDLPR(43.6606) (g) DLPR (h) CDLPR (i) DLPR (j) CDLPR

**Fig. 20.** Experimental results of DLPR and CDLPR for out-of-sample testing image 5 with high contrast and tangled patterns.

Table 11 EXPERIMENTAL RESULTS OF THE LATEST COMPETING ALGORITHMS

| Algorithm | | CGIC | TACO | LBIC |
|---|---|---|---|---|
| PSNR↑ | Original | 28.4184 | 27.6203 | 40.8254 |
| | Circular | 28.5266 | 27.6765 | 49.9477 |
| | Δ | 0.1082 | 0.0562 | 9.1224 |
| | $\Delta_m$ | **0.5037** | **0.1806** | **11.7911*** |
| | σ | 0.1187 | 0.0559 | 0.9075 |
| SSIM↑ | Original | 0.7901 | 0.7448 | 0.9949 |
| | Circular | 0.7963 | 0.7480 | 0.9989 |
| | Δ | 0.0062 | 0.0033 | 0.0040 |
| | $\Delta_m$ | 0.0249 | 0.0134 | 0.0066 |
| | σ | 0.0064 | 0.0035 | 0.0014 |
| BPSP↓ | Original | 0.5059 | 0.1503 | 1.5875 |
| | Circular | 0.5058 | 0.1473 | 1.5387 |
| | Δ | 0.0001 | 0.0030 | 0.0488 |
| | $\Delta_m$ | 0.0006 | 0.0072 | 0.0868 |
| | σ | 0.0002 | 0.0023 | 0.0172 |

## 5. Conclusion

This paper proposes the CIC framework which is a mixture of open-loop and closed-loop architectures. The open-loop structure comprises encoding and decoding units while the closed-loop structure comprises coding element, decoding element, summator, multiplier, and integrator. The proposed CIC is described by a nonlinear loop equation which is resolved by linear approximation of Taylor series expansion, and the zero steady-state error of the proposed CIC is mathematically proved. The proposed CIC can minimize the intrinsic difference between testing and training images and improve the performance of image reconstruction. The proposed CIC holds the property of Plug-and-Play and Post-Training and can be established on any existing advanced SIC algorithms. The experimental results including R-D curves on five public image compression datasets show that the proposed CIC outperforms eight open-source state-of-the-art SIC approaches. Experimental results further show that the proposed CIC is particularly effective for out-of-sample testing images with dark backgrounds, sharp edges, high contrast, grid shapes, and complex patterns.

In our future work, the proposed CIC will be verified on more public image compression datasets, such as medical and remote-sensing datasets, and more leading open-source SIC methods. The proposed CIC will also be incorporated into In-Training procedure. Some advanced control theory, such as fuzzy logic, will further be considered in the proposed CIC.

## Author Contributions

Conceptualization and Maria TROCAN, Honggui LI; Methodology, Honggui LI and Dimitri GALAYKO; Writing, Honggui LI, Sinan CHEN, and Dingtai LI; Experiment, Honggui LI, Zhengyang ZHANG, Nahid MD LOKMAN HOSSAIN, Xinfeng XU, Yinlu QIN, Ruobing WANG; Supervision, Amara AMARA and Mohamad SAWAN. All authors have read and agreed to the published version of the manuscript.

## Competing Interests

The authors declare that they have no known competing financial interests or personal relationships that could have appeared to influence the work reported in this paper.

## Acknowledgment

The authors wish to acknowledge with gratitude to all the authors of the benchmark approaches for altruistically distributing their source codes of image compression on the Microsoft GitHub website. The open-source codes allow us to easily realize the proposed algorithm depending on the benchmark approaches. The authors also wish to convey our sincere thanks to Google COLAB for its free GPU computational support.

## References

[1] S. Noy and W. Zhang, “Experimental evidence on the productivity effects of generative artificial intelligence,” *Science*, vol. 381, no. 6654, pp. 187-192, Jul. 2023, doi: 10.1126/science.adh2586.

[2] S. Jamil, M. J. Piran, M. Rahman, and O.J. Kwon, “Learning-driven lossy image compression: A comprehensive survey,” *Eng. Appl. Artif. Intell.*, vol. 123, no. B, Art. no. 106361, Aug. 2023, doi: 10.1016/j.engappai.2023.106361.

[3] C. H. Huang and J. L. Wu, “Unveiling the future of human and machine coding: A survey of end-to-end learned image compression,” *Entropy*, vol. 26, no. 5, pp. 1-35, May 2024, doi: 10.3390/e26050357.

[4] B. A. Lungisani, C. K. Lebekwe, A. M. Zungeru, and A. Yahya, “Image compression techniques in wireless sensor networks: A survey and comparison,” *IEEE Access*, vol. 10, pp. 82511-82530, Aug. 2022, doi: 10.1109/ACCESS.2022.3195891.

[5] D. Mishra, S. K. Singh, and R. K. Singh, “Deep architectures for image compression: A critical review,” *Signal Process.*, vol. 191, Art. no. 108346, Feb. 2022, doi: 10.1016/j.sigpro.2021.108346.

[6] J. Y. Li, P. G. Chen, S. Z. Yu, S. Liu, and J. Y. Jia, “MOODv2: Masked image modeling for out-of-distribution detection,” *IEEE Trans. Pattern Anal. Mach. Intell.*, vol. 46, no. 12, pp. 8994-9003, Jun. 2024, doi: 10.1109/TPAMI.2024.3412004.

[7] S. H. Li, W. R. Dai, Y. M. Fang, Z. Y. Zheng, W. Fei, H. K. Xiong, W. Zhang, “Revisiting learned image compression with statistical measurement of latent representations,” *IEEE Trans. Circuits Syst. Video Technol.*, vol. 34, no. 4, pp. 2891-2907, Apr. 2024, doi: 10.1109/TCSVT.2023.3300316.

[8] K. Tsubota and K. Aizawa, “Content-adaptive optimization framework for universal deep image compression,” *IEICE Trans Inf. Syst.,* vol. E107, no. 2, pp. 201-211, Feb. 2024, doi: 10.1587/transinf.2023EDP7114.

[9] J. M. Zhu, Y. Q. Yang, T. P. Zhang, and Z. Q. Cao, “Finite-time stability control of uncertain nonlinear systems with self-limiting control terms,” *IEEE Trans Neur. Net. Lear. Syst.*, vol. 34, no. 11, pp. 9514-9519, Nov. 2023, doi: 10.1109/TNNLS.2022.3149894.

[10] X. N. Xia, T. P. Zhang, G. P. Kang, and Y. Fang, “Adaptive control of uncertain nonlinear systems with discontinuous input and time-varying input delay,” *IEEE Trans Syst. Man Cyber-Syst.*, vol. 52, no. 11, pp. 7248-7258, Nov. 2022, doi: 10.1109/TSMC.2022.3158617.

[11] H. Li, N. M. L. Hossain, M. Trocan, M. Sawan, D. Galayko, “CMISR: circular medical image super-resolution,” *Eng. Appl. Artif. Intell.*, vol. 33, no. B, Art. no. 108222, Jul. 2024, doi: 10.1016/j.engappai.2024.108222.

[12] H. Li, M. Trocan, D. Galayko, M. Sawan, “ICRICS: iterative compensation recovery for image compressive sensing,” *Signal, Image and Video Processing*, vol. 17, no. 6, 2953-2969, Sep. 2023, doi: 10.1007/s11760-023-02516-z.

[13] H. Li, M. Trocan, M. Sawan, D. Galayko, “CSwin2SR: circular Swin2SR for compressed image super-resolution,” in IEEE 5th Int. Conf. on Artificial Intelligence Circuits and Systems (AICAS2023), Hangzhou, Zhejiang, China, 2023, pp. 1-5.

[14] P. X. Wei, Z. W. Xie, G. B. Li, and L. Lin, “Taylor neural network for real-world image super-resolution,” *IEEE Trans. Image Process.*, vol. 32, pp. 1942-1951, Apr. 2023, doi: 10.1109/TIP.2023.3255107.

[15] Y. E. Bao, W. Tan, L. F. Zheng, F. Y. Meng, W. Liu, and Y. S. Liang, “Taylor series based dual-branch transformation for learned image compression,” *Signal Process.*, vol. 212, Art. no. 109128, Nov. 2023, doi: 10.1016/j.sigpro.2023.109128.

[16] A. Ebner and M. Haltmeier, “Plug-and-Play image reconstruction is a convergent regularization method,” *IEEE Trans. Image Process.*, vol. 33, pp. 1476-1486, Mar. 2024, doi: 10.1109/TIP.2024.3361218.

[17] Y. Chen, X. F. Gui, J. S. Zeng, X. L. Zhao, and W. He, “Combining low-rank and deep Plug-and-Play priors for snapshot compressive imaging,” *IEEE Trans Neur. Net. Lear. Syst.*, vol. 35, no. 11, pp. 16396-16408, Jul. 2023, doi: 10.1109/TNNLS.2023.3294262.

[18] T. T. Wu, W. N. Wu, Y. Yang, F. L. Fan, and T. Y. Zeng, “Retinex image enhancement based on sequential decomposition with a Plug-and-Play framework,” *IEEE Trans Neur. Net. Lear. Syst.*, vol. 35, no. 10, p. 14559-14572, Jun. 2023, doi: 10.1109/TNNLS.2023.3280037.

[19] U. S. Kamilov, C. A. Bouman, G. T. Buzzard, and B. Wohlberg, “Plug-and-Play methods for integrating physical and learned models in computational imaging: theory, algorithms, and applications,” *IEEE Signal Process. Mag.*, vol. 40, no. 1, pp. 85-97, Jan. 2023, doi: 10.1109/MSP.2022.3199595.

[20] K. Zhang, Y. W. Li, W. M. Zuo, L. Zhang, G. L. Van, and R. Timofte, “Plug-and-Play image restoration with deep denoiser prior,” *IEEE Trans. Pattern Anal. Mach. Intell.*, vol. 44, no. 10, pp. 6360-6376, Nov. 2021, doi: 10.1109/TPAMI.2021.3088914.

[21] S. M. Zandavi, “Post-trained convolution networks for single image super-resolution,” *Artif. Intell.*, vol. 318, Art. no. 103882, May 2023, doi: 10.1016/j.artint.2023.103882.

[22] Y. C. Bai, X. M. Liu, K. Wang, X. Y. Ji, X. L. Wu, and W. Gao, “Deep lossy plus residual coding for lossless and near-lossless image compression,” *IEEE Trans. Pattern Anal. Mach. Intell.*, vol. 46, no. 5, pp. 3577-3594, May 2024, doi: 10.1109/TPAMI.2023.3348486.

[23] R. H. Yang and S. Mandt, “Lossy image compression with conditional diffusion models,” in the 37th Conf. on Neural Information Processing Systems, New Orleans, LA, USA, 2023, pp. 1-25.

[24] Y. C. Bai, X. Yang, X. M. Liu, J. J. Jiang, Y. W. Wang, X. Y. Ji, and W. Gao, “Towards end-to-end image compression and analysis with transformers,” in the 36th AAAI Conf. on Artificial Intelligence, Virtual, Online, 2022, pp. 104-112.

[25] Z. B. Zhang, S. Esenlik, Y. J. Wu, M. Wang, K. Zhang, and L. Zhang, “End-to-end learning-based image compression with a decoupled framework,” *IEEE Trans. Circuits Syst. Video Technol.*, vol. 34, no. 5, pp. 3067-3081, May 2024, doi: 10.1109/TCSVT.2023.3313974.

[26] D. Y. Zhang, F. Li, M. Liu, R. M. Cong, H. H. Bai, M. Wang, and Y. Zhao, “Exploring resolution fields for scalable image compression with uncertainty guidance,” *IEEE Trans. Circuits Syst. Video Technol.*, vol. 34, no. 4, pp. 2934-2948, Apr. 2024, doi: 10.1109/TCSVT.2023.3307438.

[27] N. D. Jr Guerin, R. C. da Silva, B. Macchiavello, “Learning-based image compression with parameter-adaptive rate-constrained loss,” *IEEE Signal Process. Lett.*, 31, 1099-1103, Apr. 2024, doi: 10.1109/LSP.2024.3383801.

[28] W. C. Zhang, Y. J. Liu, L. Y. Chen, J. H. Shi, X. M. Hong, and X. B. Wang, “Semantically-disentangled progressive image compression for deep space communications: Exploring the ultra-low rate regime,” *IEEE J. Sel. Area Comm*, vol. 42, no. 5, pp. 1130-1144, May 2024, doi: 10.1109/JSAC.2024.3369654.

[29] H. S. Fu, F. Liang, J. P. Lin, B. Li, M. Akbari, J. Liang, G. H. Zhang, D. Liu, C. J. Tu, and J. N. Han, “Learned image compression with Gaussian-Laplacian-logistic mixture model and concatenated residual modules,” *IEEE Trans. Image Process.*, vol. 32, pp. 2063-2076, May 2023, doi: 10.1109/TIP.2023.3263099.

[30] Z. H. Duan, M. Lu, Z. Ma, and F. Q. Zhu, “Lossy image compression with quantized hierarchical VAEs,” in the IEEE Winter Conf. on Applications of Computer Vision, Waikoloa, HI, USA, 2023, pp. 198-207.

[31] A. Timur, “Vector quantized variational image compression,” Skolkovo Institute of Science and Technology, Moscow, Russia, Tech. Rep. of Bayesian Methods of Machine Learning, Sep. 2022.

[32] S. L. Cai, L. Q. Chen, Z. J. Zhang, X. Y. Zhao, J. H. Zhou, Y. X. Peng, L. X. Yan, S. Zhong, and X. Zou, “I2C: Invertible continuous codec for high-fidelity variable-rate image compression,” *IEEE Trans. Pattern Anal. Mach. Intell.*, vol. 46, no. 6, pp. 4262-4279, Jun. 2024, doi: 10.1109/TPAMI.2024.3356557.

[33] H. S. Fu, F. Liang, J. Liang, B. L. Li, G. H. Zhang, and J. N. Han, “Asymmetric learned image compression with multi-scale residual block, importance scaling, and post-quantization filtering,” *IEEE Trans. Circuits Syst. Video Technol.*, vol. 33, no. 8, pp. 4309-4321, Aug. 2023, doi: 10.1109/TCSVT.2023.3237274.

[34] G. Zhang, X. F. Zhang, and L. Tang, "Enhanced quantified local implicit neural representation for image compression," *IEEE Signal Process. Lett.*, vol. 30, pp. 1742-1746, Mar. 2023, doi: 10.1109/LSP.2023.3334956.

[35] J. Y. Guo, D. Xu, and G. Lu, "CBANet: Toward complexity and bitrate adaptive deep image compression using a single network," *IEEE Trans. Image Process.*, vol. 32, pp. 2049-2062, May 2023, doi: 10.1109/TIP.2023.3251020.

[36] Z. Y. Jiang, X. H. Liu, A. N. Li, and G. Y. Wang, "Enhancing high-resolution image compression through local-global joint attention mechanism," *IEEE Signal Process. Lett.*, vol. 31, pp. 1044-1048, Jun. 2024, doi: 10.1109/LSP.2024.3383963.

[37] B. Li, Y. J. Li, J. C. Luo, X. R. Zhang, C. Y. Li, Z. M. Chenjin, and Y. Liang, "Learned image compression via neighborhood-based attention optimization and context modeling with multi-scale guiding," *Eng. Appl. Artif. Intell.*, vol. 129, Mar. 2024, Art. no. 107596, doi: 10.1016/j.engappai.2023.107596.

[38] Z. S. Tang, H. L. Wang, X. K. Yi, Y. Zhang, S. Kwong, and C. C. J. Kuo, "Joint graph attention and asymmetric convolutional neural network for deep image compression," *IEEE Trans. Circuits Syst. Video Technol.*, vol. 33, no. 1, pp. 421-433, Jan. 2023, doi: 10.1109/TCSVT.2022.3199472.

[39] J. Q. Shi, M. Lu, and Z. Ma, "Rate-distortion optimized post-training quantization for learned image compression," *IEEE Trans. Circuits Syst. Video Technol.*, vol. 34 no. 5, pp. 3082-3095, May 2024, doi: 10.1109/TCSVT.2023.3323015.

[40] Z. H. Duan, M. Lu, J. Ma, Y. N. Huang, Z. Ma, and F. Q. Zhu, "QARV: Quantization-aware ResNet VAE for lossy image compression," *IEEE Trans. Pattern Anal. Mach. Intell.*, vol. 46, no. 1, pp. 436-450, Jan. 2024, doi: 10.1109/TPAMI.2023.3322904.

[41] S. H. Li, H. Li, W. R. Dai, C. L. Li, J. N. Zou, and H. K. Xiong, "Learned progressive image compression with dead-zone quantizers," *IEEE Trans. Circuits Syst. Video Technol.*, vol. 33, no. 6, pp. 2962-2978, Jun. 2023, doi: 10.1109/TCSVT.2022.3229701.

[42] H. Son, T. Kim, H. Lee, and S. Lee, "Enhanced standard compatible image compression framework based on auxiliary codec networks," *IEEE Trans. Image Process.*, vol. 31, pp. 664-677, Jan. 2022, doi: 10.1109/TIP.2021.3134473.

[43] J. F. Li, X. Y. Liu, Y. Q. Gao, L. Zhuo, and J. Zhang, "BARRN: A blind image compression artifact reduction network for industrial IoT systems," *IEEE Trans Ind. Inform.*, vol. 19, no. 9, pp. 9479-9490, Sep. 2023, doi: 10.1109/TII.2022.3228777.

[44] L. Ma, Y. F. Zhao, P. X. Peng, and Y. H. Tian, "Sensitivity decouple learning for image compression artifacts reduction," *IEEE Trans. Image Process.*, vol. 33, pp. 3620-3633, Jun. 2024, doi: 10.1109/TIP.2024.3403034.

[45] J. H. Hu, G. X. Luo, B. Wang, W. M. Wu, J. H. Yang, and J. D. Guo, "Residual network for image compression artifact reduction," *Int. J. Pattern Recogn. Artif. Intell.*, vol. 38, no. 2, Mar. 2024, Art. no. 2454001, doi: 10.1142/S0218001424540016.

[46] H. G. Chen, X. H. He, H. Yang, L. B. Qing, and Q. Z. Teng, "A feature-enriched deep convolutional neural network for JPEG image compression artifacts reduction and its applications," *IEEE Trans Neur. Net. Lear. Syst.*, vol. 33, no. 1, pp. 430-444, Jan. 2022, doi: 10.1109/TNNLS.2021.3124370.

[47] Anqi Li, Yuxi Liu, Huihui Bai, Feng Li, Runmin Cong, Meng Wang, and Yao Zhao, "Once-for-all: controllable generative image compression with dynamic granularity adaption," in the 13th International Conference on Learning Representations (ICLR2025), EXPO, Singapore, 2025, pp. 1-22.

[48] H. Lee, M. Kim, J.-H. Kim, S. Kim, D. Oh, and J. Lee, "Neural image compression with text-guided encoding for both pixel-level and perceptual fidelity," in the 41st International Conference on Machine Learning (ICML2024), Vienna, Austria, 2024, pp. 26715-26730.

[49] F. Kamisli, "End-to-end learned block-based image compression with block-level masked convolutions and asymptotic closed-loop training," *Multimed. Tools Appl.*, online, pp. 1-23, Sep. 2024, doi: 10.1007/s11042-024-20315-7.

# BIOGRAPHIES

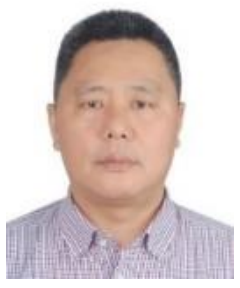

**Honggui LI** received a B.S. degree in electronic science and technology from Yangzhou University and received a Ph.D. degree in mechatronic engineering from Nanjing University of Science and Technology. He is a senior member of the Chinese Institute of Electronics. He is a visiting scholar and a post-doctoral fellow at Institut Supérieur d'Électronique de Paris for one year. He is an associate professor of electronic science and technology and a postgraduate supervisor of electronic science and technology at Yangzhou University. He is a reviewer for some international journals and conferences. He is the author of over 30 refereed journal and conference articles. His current research interests include image processing, machine learning, and integrated circuits engineering.

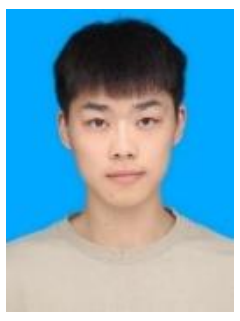

**Sinan CHEN** received a B.E. degree in electronics information engineering from Yangzhou University in China. He is now studying for a master's degree in integrated circuits engineering at Yangzhou University in China. His research interests include image compressive sensing, machine learning, and integrated circuits engineering.

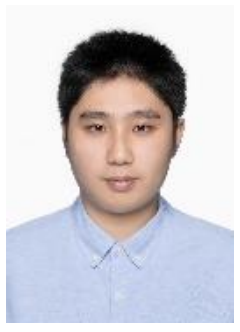

**Dingtai LI** received a B.M. degree in rehabilitation of Chinese medicine from Nanjing University of Chinese Medicine in China. He is now studying for a master's degree in acupuncture and massage at Shanghai University of Traditional Chinese Medicine in China. His research interests include Qigong, acupuncture, massage, image processing, and machine learning.

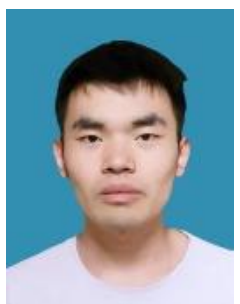

**Zhengyang ZHANG** received a B.E. degree in electronics information engineering from Yangzhou University in China. He is now studying for a master's degree in communication engineering at Yangzhou University in China. His research interests include image super-resolution, machine learning, and communication engineering.

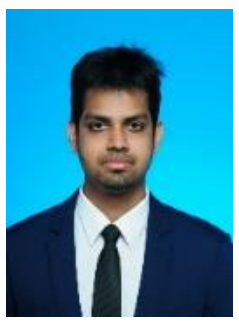

**Nahid MD LOKMAN HOSSAIN** received a B.S. degree in computer science and technology from Chongqing University of Posts and Telecommunications in China. He is now studying for a master's degree in software engineering at Yangzhou University in China. His research interests include image super-resolution, machine learning, innovation software, cloud computing, and big data.

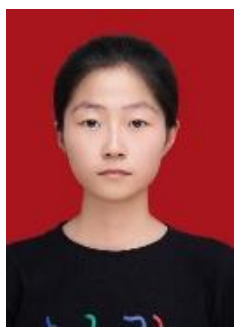

**Xinfeng XU** is now studying for a B.E. degree in electronic information engineering at Yangzhou University in China. Her research interests include image processing and machine learning.

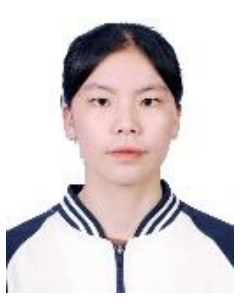

**Yinlu QIN** is now studying for a B.E. degree in electronic information engineering at Yangzhou University in China. Her research interests include image processing and machine learning.

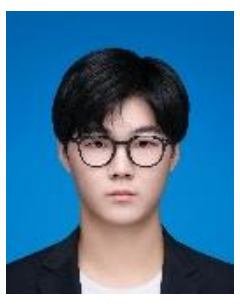

**Ruobing WANG** is now studying for a B.E. degree in electronic information engineering at Yangzhou University in China. His research interests include image processing and machine learning.

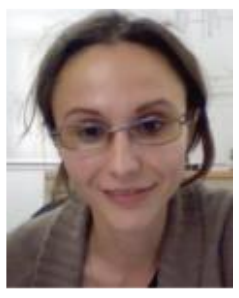

**Maria TROCAN** received a M.Eng. in Electrical Engineering and Computer Science from the Politehnica University of Bucharest, a Ph.D. in Signal and Image Processing from Telecom ParisTech, and the Habilitation to Lead Researches (HDR) from Pierre & Marie Curie University (Paris 6). She has joined Joost - Netherlands, where she worked as a research engineer involved in the design and development of video transcoding systems. She is firstly Associate Professor, then Professor at Institut Superieur d'Electronique de Paris (ISEP). She is an Associate Editor for the Springer Journal on Signal, Image and Video Processing and a Guest Editor for several journals (Analog Integrated Circuits and Signal Processing, IEEE Communications Magazine, etc.). She is an active member of IEEE France and served as a counselor for the ISEP IEEE Student Branch, IEEE France Vice-President responsible for Student Activities, and IEEE Circuits and Systems Board of Governors member, as Young Professionals representative. Her current research interests focus on image and video analysis & compression, sparse signal representations, machine learning, and fuzzy inference.

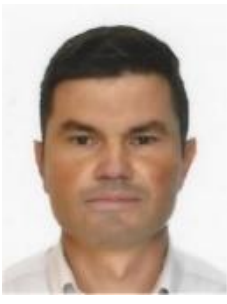

**Dimitri GALAYKO** received a bachelor's degree from Odessa State Polytechnic University in Ukraine, a master's degree from the Institute of Applied Sciences of Lyon in France, and a Ph.D. degree from University of Lille in France. He made his Ph.D. thesis at the Institute of Microelectronics and Nanotechnologies. His Ph.D. dissertation was on the design of micro-electromechanical silicon filters and resonators for radio communications. He is a Professor at the LIP6 research laboratory of Sorbonne University in France. His research interests include the study, modeling, and design of nonlinear integrated circuits for sensor interfaces and mixed-signal applications. His research interests also include machine learning and fuzzy computing.

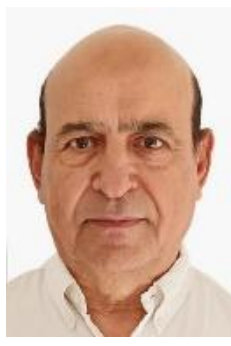

**Amara AMARA** (Senior Member, IEEE) received the HDR (Confirmation of Leading Research Capabilities) degree from Evry University and the Ph.D. degree from Paris VI University in 1989. In 1988, he joined the IBM Research and Development Laboratory, Corbeil-Essonnes, as a Visiting Researcher, where he was involved in SRAM memory design with advanced CMOS technologies. In 1992, he joined the Paris Institute for Electronics (ISEP) in charge of the Microelectronics Laboratory, where he headed a joint team (Paris VI and ISEP) involved in highspeed GaAs VLSI circuit design. He established the LISTE Laboratory composed of more than 40 researchers in the fields of micro and nano electronics, image and signal processing, and big data processing and analysis. He is the coauthor of three books on Molecular Electronics, Double Gate Devices and Circuits, and Emerging Technologies. He is the author or coauthor of more than 100 conference papers and journal articles. He was also the advisor of more than 20 Ph.D. students. He was the Deputy Managing Director of ISEP in charge of Research and International Cooperation since March 2017. He then joined in June 2017 "Terre des hommes" (Tdh) an international NGO specialized in child protection where he was in charge of ICT for Development (ICT4D) and Artificial Intelligence for Health. He was the President of the CAS Society from 2020 to 2021. He is the former IEEE France Section Chair and the Co-Founder and the Chair of the IEEE France CASS Chapter.

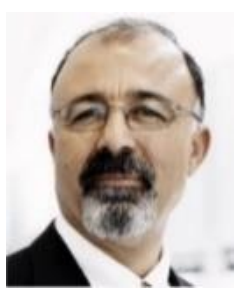

**Mohamad SAWAN** received a Ph.D. degree in electrical engineering from the University of Sherbrooke, Sherbrooke, QC, Canada, in 1990. He was a Chair Professor awarded with the Canada Research Chair in Smart Medical Devices (2001–2015) and was leading the Microsystems Strategic Alliance of Quebec - ReSMiQ (1999–2018). He is a Professor of Microelectronics and Biomedical Engineering, in leave of absence from Polytechnique Montréal, Canada. He joined Westlake University, Hangzhou, China, in January 2019, where he is a Chair Professor, Founder, and Director of the Center for Biomedical Research And Innovation (CenBRAIN). He has published more than 800 peer-reviewed articles, two books, ten book chapters, and 12 patents. He founded and chaired the IEEE-Solid State Circuits Society Montreal Chapter (1999–2018) and founded the Polystim Neurotech Laboratory, Polytechnique

Montréal (1994–present), including two major research infrastructures intended to build advanced Medical devices. He is the Founder of the International IEEE-NEWCAS Conference, and the Co-Founder of the International IEEE-BioCAS, ICECS, and LSC conferences. He is a Fellow of the Royal Society of Canada, a Fellow of the Canadian Academy of Engineering, and a Fellow of the Engineering Institutes of Canada. He is also the "Officer" of the National Order of Quebec. He has served as a member of the Board of Governors (2014–2018). He is the Vice-President of Publications (2019–present) of the IEEE CAS Society. He received several awards, among them the Queen Elizabeth II Golden Jubilee Medal, the Barbara Turnbull 2003 Award for spinal cord research, the Bombardier and Jacques-Rousseau Awards for academic achievements, the Shanghai International Collaboration Award, and the Medal of Merit from the President of Lebanon for his outstanding contributions. He was the Deputy Editor-in-Chief of the IEEE TRANSACTIONS ON CIRCUITS AND SYSTEMS-II: EXPRESS BRIEFS (2010–2013); the Co-Founder, an Associate Editor, and the Editor-in-Chief of the IEEE TRANSACTIONS ON BIOMEDICAL CIRCUITS AND SYSTEMS; an Associate Editor of the IEEE TRANSACTIONS ON BIOMEDICALS ENGINEERING; and the International Journal of Circuit Theory and Applications.